\documentclass[twocolumn,trackchanges]{aastex631}

\usepackage{amsmath}
\usepackage{bm}
\usepackage{graphicx}
\usepackage{multirow}

\begin{document}

\title{Trace the Accretion Geometry of H 1743--322 with Type C Quasi-periodic Oscillations in Multiple Outbursts}

\correspondingauthor{Qing C. Shui}
\email{shuiqc@ihep.ac.cn}
\correspondingauthor{S. Zhang}
\email{szhang@ihep.ac.cn}
\correspondingauthor{Yu P. Chen}
\email{chenyp@ihep.ac.cn}

\author[0000-0001-5160-3344]{Qing C. Shui}
\affiliation{Key Laboratory of Particle Astrophysics, Institute of High Energy Physics, Chinese Academy of Sciences, 100049, Beijing, China}
\affiliation{University of Chinese Academy of Sciences, Chinese Academy of Sciences, 100049, Beijing, China}

\author{S. Zhang}
\affiliation{Key Laboratory of Particle Astrophysics, Institute of High Energy Physics, Chinese Academy of Sciences, 100049, Beijing, China}


\author{Yu P. Chen}
\affiliation{Key Laboratory of Particle Astrophysics, Institute of High Energy Physics, Chinese Academy of Sciences, 100049, Beijing, China}

\author[0000-0001-5586-1017]{Shuang N. Zhang}
\affiliation{Key Laboratory of Particle Astrophysics, Institute of High Energy Physics, Chinese Academy of Sciences, 100049, Beijing, China}
\affiliation{University of Chinese Academy of Sciences, Chinese Academy of Sciences, 100049, Beijing, China}

\author[0000-0003-3188-9079]{Ling D. Kong}
\affiliation{Key Laboratory of Particle Astrophysics, Institute of High Energy Physics, Chinese Academy of Sciences, 100049, Beijing, China}
\affiliation{Institut f\"{u}r Astronomie und Astrophysik, Kepler Center for Astro and Particle Physics, Eberhard Karls, Universit\"{a}t, Sand 1, D-72076 T\"{u}bingen, Germany}

\author[0000-0002-6454-9540]{Peng J. Wang}
\affiliation{Key Laboratory of Particle Astrophysics, Institute of High Energy Physics, Chinese Academy of Sciences, 100049, Beijing, China}
\affiliation{University of Chinese Academy of Sciences, Chinese Academy of Sciences, 100049, Beijing, China}

\author{L. Ji}
\affiliation{School of Physics and Astronomy, Sun Yat-Sen University, Zhuhai, 519082, China}

\author[0000-0002-0638-088X]{Hong X. Yin}
\affiliation{Shandong Key Laboratory of Optical Astronomy and Solar-Terrestrial Environment, School of Space Science and Physics, Institute of Space Sciences, Shandong University, Weihai, Shandong 264209, China}

\author[0000-0002-9796-2585]{Jin L. Qu}
\affiliation{Key Laboratory of Particle Astrophysics, Institute of High Energy Physics, Chinese Academy of Sciences, 100049, Beijing, China}

\author[0000-0002-2705-4338]{L. Tao}
\affiliation{Key Laboratory of Particle Astrophysics, Institute of High Energy Physics, Chinese Academy of Sciences, 100049, Beijing, China}

\author[0000-0002-2749-6638]{Ming Y. Ge}
\affiliation{Key Laboratory of Particle Astrophysics, Institute of High Energy Physics, Chinese Academy of Sciences, 100049, Beijing, China}

\author[0000-0002-5554-1088]{Jing Q. Peng}
\affiliation{Key Laboratory of Particle Astrophysics, Institute of High Energy Physics, Chinese Academy of Sciences, 100049, Beijing, China}
\affiliation{University of Chinese Academy of Sciences, Chinese Academy of Sciences, 100049, Beijing, China}

\author[0000-0003-4856-2275]{Z. Chang}
\affiliation{Key Laboratory of Particle Astrophysics, Institute of High Energy Physics, Chinese Academy of Sciences, 100049, Beijing, China}

\author{J. Li}
\affiliation{CAS Key Laboratory for Research in Galaxies and Cosmology, Department of Astronomy, University of Science and Technology of China, Hefei 230026, China}
\affiliation{School of Astronomy and Space Science, University of Science and Technology of China, Hefei 230026, China}

\author{P. Zhang}
\affiliation{College of Science, China Three Gorges University, Yichang 443002, China}



\begin{abstract}
We present a systematic analysis of type C quasi-periodic oscillation (QPO) observations of H 1743--322 throughout the \emph{Rossi X-ray Timing Explorer} (\emph{RXTE}) era. We find that, while different outbursts have significant flux differences, they show consistent positive correlations between the QPO fractional root-mean-square (rms) amplitude and non-thermal fraction of the emission, which indicate an independence of the intrinsic QPO rms on individual outburst brightness in H 1743--322. However, the dependence of the QPO rms on frequency is different between the outburst rise and decay phases, where QPO fractional rms of the decay phase is significantly lower than that of the rise phase at low frequencies. The spectral analysis also reveals different ranges of coronal temperature between the two outburst stages. A semi-quantitative analysis shows that the Lense-Thirring precession model could be responsible for the QPO rms differences, requiring a variable coronal geometric shape. However, the variable-Comptonization model could also account for the findings. The fact that the rms differences and the hysteresis traces in the hardness-intensity diagram (HID) accompany each other indicates a connection between the two phenomena. By correlating the findings with QPO phase lags and the quasi-simultaneous radio flux previously published, we propose there could be corona-jet transitions in H 1743--322 similar to those that have been recently reported in GRS 1915+105.

\end{abstract}



\section{Introduction} \label{sec:intro}

Undergoing outbursts occasionally after staying with faint luminosity for a long time in quiescence is the primary feature of low mass black hole X-ray binaries (BHXRBs) \citep{2006ARA&A..44...49R,2007A&ARv..15....1D}. Most complete outbursts are observed to have four 
typical states: the low/hard state (LHS), hard intermediate state (HIMS), soft intermediate state (SIMS) and high soft state (HSS), characterized by different X-ray spectral and variability properties \citep[][]{2005A&A...440..207B,2009MNRAS.396.1370F}. In the outburst rise phase, the system starts from quiescence, increases luminosity with strong variability and non-thermal dominated spectra in the LHS, experiences the hard-to-soft transition in the HIMS and SIMS, and then stays in the HSS with the thermal dominated spectra and weakest variability for weeks. With the decreasing accretion rate, the system goes through the soft-to-hard transition back to the LHS, and fades in quiescence. The different state transition luminosity between the rise and decay phases directly leads to the trace of canonical `q' shape in the hardness-intensity diagram \citep[HID,][]{,2001ApJS..132..377H,2005Ap&SS.300..107H}, and the so-called hysteresis effect which is still not well understood in BHXRBs \citep{2003A&A...409..697M,2004MNRAS.351..791Z,2021ApJ...915L..15W}. In addition to these features in the X-ray energy bands, BHXRBs are also characterized by radio/infrared emission, which is  generally believed to be associated with relativistic jets \citep[][and references therein]{2001MNRAS.322...31F,2004ARA&A..42..317F,2022NatAs...6..577M}. In the hard state, radio emission with a flat spectrum ($S_{\nu}\propto\nu^\alpha$ with $\alpha\sim0$) is interpreted as self-absorbed synchrotron emission from an optically thick, steady and compact jet \citep[][]{1979ApJ...232...34B,2001MNRAS.322...31F}.During the hard-to-soft transition, the steady and compact jet is gradually quenched, where the radio emission, if present, is thought to be attributed to optically thin synchrotron emission from transient ejected plasma clouds with relativistic speeds ($v\sim c$) \citep[][]{2004MNRAS.355.1105F}.

The X-ray spectrum of a BHXRB usually consists of a thermal and a non-thermal component, the former is believed to come from a geometrically thin and optically thick disc \citep{1973A&A....24..337S,1974MNRAS.168..603L}, while the latter is produced by the Comptonization of soft photons. However, the geometry of the Comptonizing medium is relatively less clear, which could be an extended cloud consisting of hot electrons ($\sim 100$ keV) called ``corona'' \citep{1976SvAL....2..191B,1980A&A....86..121S,1994ApJ...434..570T,1996MNRAS.283..193Z,1999MNRAS.303L..11Z,1999MNRAS.309..561Z} or/and the jet-base \citep{2005ApJ...635.1203M,2021NatCo..12.1025Y}. A portion of Comptonized photons can irradiate the disc and then end up as the reflection component which has abundant features, like broad emission lines and Compton hump, etc. \citep[][and references therein]{2010MNRAS.409.1534D,2014ApJ...782...76G,2015ApJ...813...84G}.  

Low frequency quasi-periodic oscillations \citep[LFQPOs, roughly 0.1--30 Hz,][]{1989ARA&A..27..517V} are the most prominent features observed in the power density spectrum (PDS) of BHXRBs, with the classification of type A, B and C based on the centroid frequency, quality factor and root-mean-square (rms) amplitude \citep{1999ApJ...526L..33W,2005ApJ...629..403C,2006ARA&A..44...49R}. The appearance of type C QPOs is frequent in the LHS and HIMS with strong amplitudes (fractional rms $\sim10\%$) and flat-top noise components in the PDS. In the past few decades, several models has been proposed to explain the dynamical origin of QPOs based either on the geometric or the intrinsic properties of the accretion flow. Some examples of the \emph{intrinsic} models are trapped corrugation modes \citep{,1990PASJ...42...99K,1999PhR...311..259W}, the Accretion-ejection instability model \citep[AEI;][]{1999A&A...349.1003T} and the Two-Component Advection Flow model \citep[TCAF;][]{1996ApJ...457..805M}, etc.. For the \emph{geometric} models, most of them are related to the relativistic Lense-Thirring (L-T) precession, which was originally invoked to be the dynamic mechanism of QPOs by \citet{1998ApJ...492L..59S}. As an extension of the relativistic precession model \citep[RPM;][]{1999ApJ...524L..63S}, the L-T precession model proposed by \citet{2009MNRAS.397L.101I} assumes the entire hot flow precesses within the inner radius of the truncated disc \citep{1997ApJ...489..865E}. We refer readers to \citet{2019NewAR..8501524I} for recent reviews of observations and theories of LFQPOs.

In addition to the frequency, the radiative properties of QPOs, e.g. rms amplitudes and time lags, etc.,  also provide extra useful information. Since several observational studies have found that, for most type C QPOs, the variability increases with the photon energy \citep{2017ApJ...845..143Z,2018ApJ...866..122H,2020JHEAp..25...29K,2020MNRAS.494.1375Z} and no prominent disc-like component exists in the rms spectra \citep{2006MNRAS.370..405S,2013MNRAS.431.1987A,2016MNRAS.458.1778A}, the radiative mechanism of the type C QPO should be strongly related to the Comptonized emission. The inclination dependence of amplitudes and time lags \citep[see][]{2015MNRAS.447.2059M,2017MNRAS.464.2643V} and reflection variability extracted from the phase-resolved spectroscopy \citep[see][]{2015MNRAS.446.3516I,2016MNRAS.461.1967I,2017MNRAS.464.2979I} add support to a geometrical origin, especially the L-T precession model \citep{2009MNRAS.397L.101I}. However, these observational findings from different sources show very diverse lacking quantitative explanations, and most recently, \citet[][]{2022MNRAS.511..255N} applied the L-T precession model to fit the phase-resolved spectroscopy of GRS 1915+105 and found an unexpectedly long thermalization time-scale of $\sim70$ ms, which is incompatible with the soft lags ($\sim1$ ms) found in other BHXRBs, e.g. MAXI J1820+070 \citep[][]{2019Natur.565..198K} and GX 339--4 \citep[][]{2011MNRAS.414L..60U}. These inconsistent findings indicate L-T precession model, especially its radiative part, could be incomplete. \citet{2020MNRAS.492.1399K} have recently proposed a time-dependent Comptonization model following the work of \citet{1998MNRAS.299..479L} and \citet{2014MNRAS.445.2818K} to quantitatively explain the energy-dependence of phase lags and rms amplitudes of kilohertz QPOs in neutron-star systems. This model has been successfully applied to BHXRBs, e.g. GRS 1915+105 and MAXI J1348--630 \citep[see][]{2021MNRAS.503.5522K,2021MNRAS.501.3173G,2022NatAs...6..577M,2022MNRAS.513.4196G}. The updated version of the model \citep{2022MNRAS.515.2099B}, incorporating a disc-blackbody as the seed-photon source, has been applied to fit the rms amplitude and phase lag spectra of type C QPOs in MAXI J1535--571, using \emph{Insight}-HXMT data in the 1--100 keV energy range \citep[][]{2022MNRAS.512.2686Z}.

The transient BHXRB H 1743--322 was first observed by \emph{Ariel}-\uppercase\expandafter{\romannumeral5} satellite in 1977 August \citep{1977IAUC.3099....3K}, located at $\rm RA=17^h46^m15^s.596$ and $\rm Dec=-32^{\circ}14'00''.860$. Based on the X-ray/radio observations of the two-sided jet in the 2003 outburst, \citet{2012ApJ...745L...7S} determined a distance of $8.5\pm0.8\ \rm kpc$, and an inclination angle of $75^{\circ}\pm3^{\circ}$, respectively. By applying the relativistic accretion disc model in the spectral fitting, they also estimated a black hole spin of $a_*=0.2\pm0.3$. H 1743--322 is an active black hole transient source and was monitored by \emph{RXTE} to undergo outbursts frequently between 2003 and 2011. The 2003 outburst is the brightest one with observations in multiple wavelengths \citep[see][]{2003A&A...411L.421P,2005ApJ...623..383H,2009ApJ...698.1398M}, then two much fainter outbursts in 2004 and 2005, respectively, followed behind \citep{2017MNRAS.466.1372B,2011MNRAS.414..677C}. In 2008, two outbursts (2008a and 2008b) were observed in both of X-ray and radio bands, but the 2008a outburst is classified as failed-transition outburst for the short outburst cycle without experiencing soft states \citep{2009MNRAS.398.1194C,2010MNRAS.401.1255J,2011MNRAS.414..677C}. Then this source entered into a new outburst during 2009 \citep{2010A&A...522A..99C,2010MNRAS.408.1796M}, and exhibited the last three outbursts in 2010 (2010a and 2010b) and 2011 \citep{2013MNRAS.431.2285Z}. Recently, \citet{2020A&A...637A..47A} systematically analysed the spectral evolution of the outbursts of H 1743--322 in the 
\emph{RXTE} era. In the post \emph{RXTE} era, there were 2012, 2014, 2016 and 2018 outburst monitored by other instruments \citep{2014ApJ...789..100S,2016MNRAS.460.1946S,2020ApJ...893..142C,2020MNRAS.491L..29W,2022MNRAS.512.4541W}.

In this work, we systematically investigate type C QPO data born out of \emph{RXTE} observations in both of the rise and decay phases of seven outbursts from H 1743--322. We introduce observations and data reductions in Section~\ref{sec:2}, present the data analysis and results in Section~\ref{sec:3} , discuss these in Section~\ref{sec:4} and finally summarize in Section~\ref{sec:5}.

\section{Observations and Data reductions} \label{sec:2}

\begin{figure*}
\centering
\includegraphics[width=17cm]{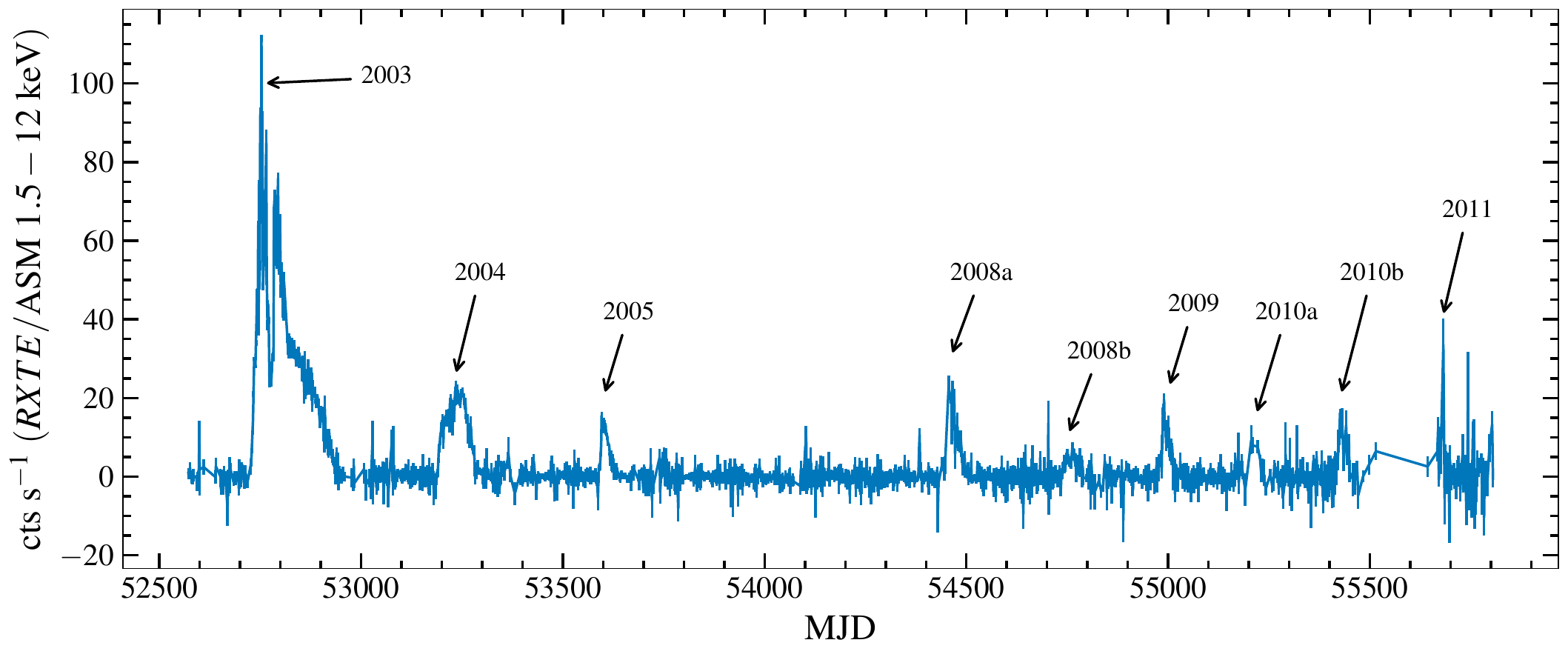}
    \caption{The \emph{RXTE}/ASM light curve of H 1743--322 in the 1.5--12 keV energy band with a resolution of one point per day. In the \emph{RXTE} era, there are nine outbursts between 2003 and 2011, where we focus on the 2003, 2008a, 2008b, 2009, 2010a, 2010b and 2011 outburst which contain abundant type C QPO observations.}
    \label{fig:1}
\end{figure*}

Fig.~\ref{fig:1} shows the \emph{RXTE}/All Sky Monitor (ASM) light-curve of H 1743--322 in the 1.5--12 keV energy band. 
Between 2003 and 2011, this source experienced nine outbursts, and we focus on the 2003, 2008a, 2008b, 2009, 2010a, 2010b and 2011 outburst which contain abundant type C QPO observations carried out by \emph{RXTE}. 
The \verb'HEASOFT' software package version 6.28 is used for the data analysis. We generate good time internals (GTIs) with the constraints that the elevation angle (ELV) is larger than $10^{\circ}$ and angular distance between the pointing position and source (OFFSET) is less than $0.02^{\circ}$. 
The standard data products of e.g. light curve, spectrum and background are produced from observational data of Proportional Counter Array (PCA). 
Since combining large spectral data sets with different PCA configurations can produce large systematic errors \citep[see][]{2009ApJ...693.1621S}, we only use data from Proportional Counter Unit (PCU) 2 in our subsequent spectral analysis because it is the only unit which was 100\% on during the observations. 
Data of standard 2 mode are adopted for the spectral analysis in the energy range of 3--30 keV without any groupings and binnings.
To account for calibration uncertainties, a systematic error of 0.5 per cent is added to spectral fittings. 
We generate the power density spectrum (PDS) using the light curve born out of `Event' and `Binned' modes in the energy range of 3--30 keV (PCA channels 7--71 of the calibration epoch 5\footnote{\url{https://heasarc.gsfc.nasa.gov/docs/xte/e-c_table.html}}), which is the same as that of the spetral fittings. 
Then both of the spectral and timing analyses are performed using XSPEC version 12.12.0.

\section{Analysis and results} \label{sec:floats}

\label{sec:3}
\subsection{Power Density Spectra}
\label{sec:3.1}
\begin{figure*}
\centering
\begin{minipage}[c]{0.49\textwidth}
\centering
    \includegraphics[width=\linewidth]{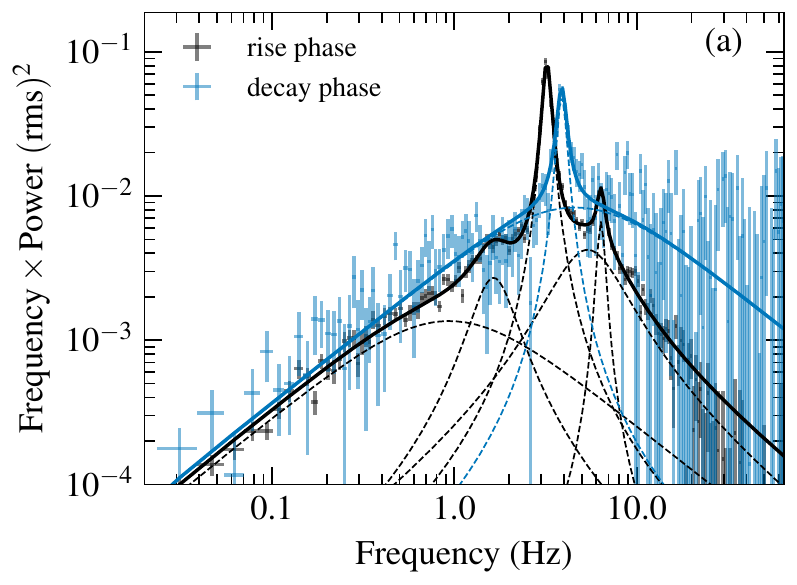}
\end{minipage}%
\begin{minipage}[c]{0.49\textwidth}
\centering
    \includegraphics[width=\linewidth]{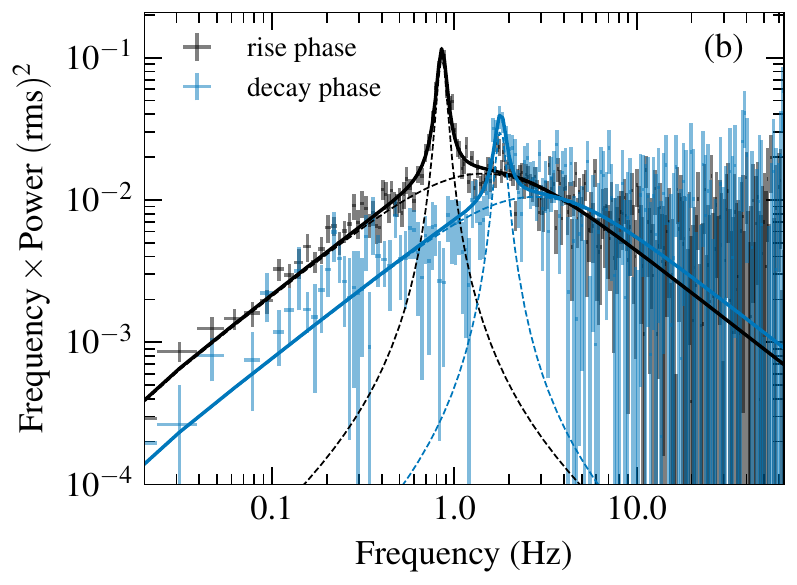}
\end{minipage}
\caption{Representative power density spectra from the outburst rise (plotted in black) and decay phase (plotted in blue). Power density spectra are produced in the 3--30 keV energy range and 1/64--64 Hz frequency range, then fitted with the model consisting of multiple Lorentzian functions (dotted lines). For panel (a), strong QPO components appear at high frequencies ($f_{\rm QPO} > 2$ Hz), where QPO amplitudes of the two outburst phases are comparable. For panel (b), the QPO components appear at low frequencies ($f_{\rm QPO} < 2$ Hz), where QPO amplitude of the rise phase is significantly larger than that of the decay phase.}
\label{fig:2}
\end{figure*}

For each observation in the present study, we produce the PDS in the 1/64–-64 Hz frequency range with the 8-ms time resolution by taking Miyamoto normalization \citep[][]{1990A&A...230..103B,1991ApJ...383..784M}. All the PDS are fitted with several Lorentzian functions (see Fig.~\ref{fig:2}). In the PDS fittings, we use at least two Lorentzian functions to fit the broad band noise and QPO signal, respectively. The centriod frequency of the Lorentzian fitted the low-frequency broad band noise is fixed at zero. However, if there are any other significant residual structures (e.g. the second QPO harmonic), more Lorentzian functions would be added to the fitting model. Since we primarily focus on the QPO signal, whether to add more Lorentzians depends on whether the Lorentzian functions used to fit the QPO harmonics describe the QPO components well in the case of the current total model. To do this, after fitting with the current model, we retain the best-fit parameters and remove the Lorentzian components fitted the QPO harmonics to check whether the residual structures from the data/model ratio plots are Lorentzian-like shapes and broad band noises are well fitted.  Following \citet{2015MNRAS.447.2059M}, if a QPO consists of multiple harmonic peaks, the QPO fractional rms is computed by adding in quadrature the rms of the harmonic peaks (i.e. QPO fractional rms is computed by $\sqrt{\sum{P_{i}}}$, where $P_{i}$ is the power calculated with integration of the $i$-th QPO harmonic Lorentzian function). Additionally, considering the background contribution, the QPO fractional rms reported in the present study is finally computed by 
\begin{equation}
\label{eq:1}
{\rm rms}=\sqrt{\sum_{i}{P_{i}}}\times[(S+B)/S],     
\end{equation}
where $S$ and $B$ are the source and background average rate, respectively \citep[see][]{2015ApJ...799....2B}. The best-fit QPO parameters of the present study are presented in Appendix \ref{sec:A4}.

\subsection{Energy Spectra}
\label{sec:3.2}
\begin{figure*}
\centering
\begin{minipage}[c]{0.49\textwidth}
\centering
    \includegraphics[width=\linewidth]{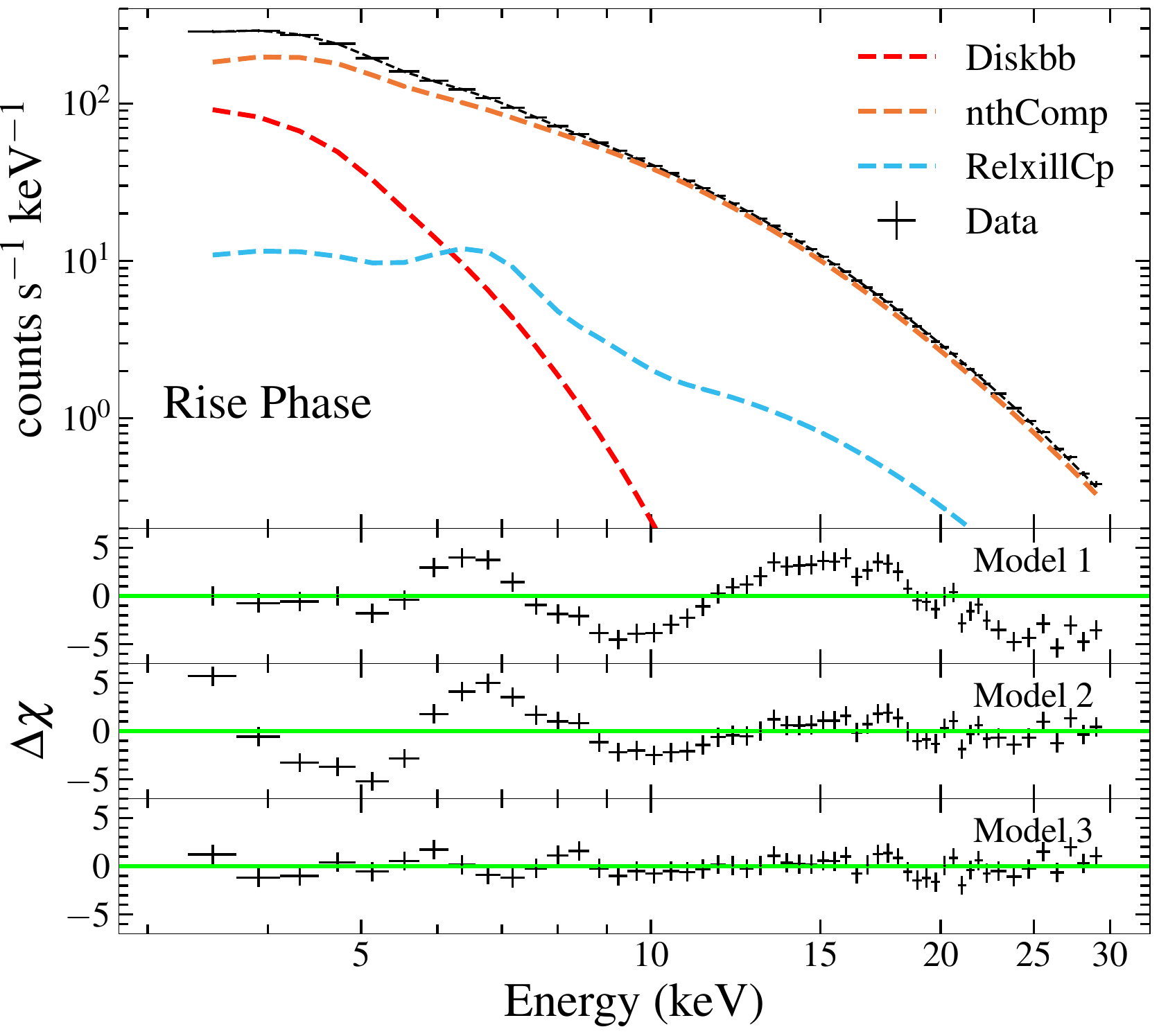}
\end{minipage}%
\begin{minipage}[c]{0.5\textwidth}
\centering
    \includegraphics[width=\linewidth]{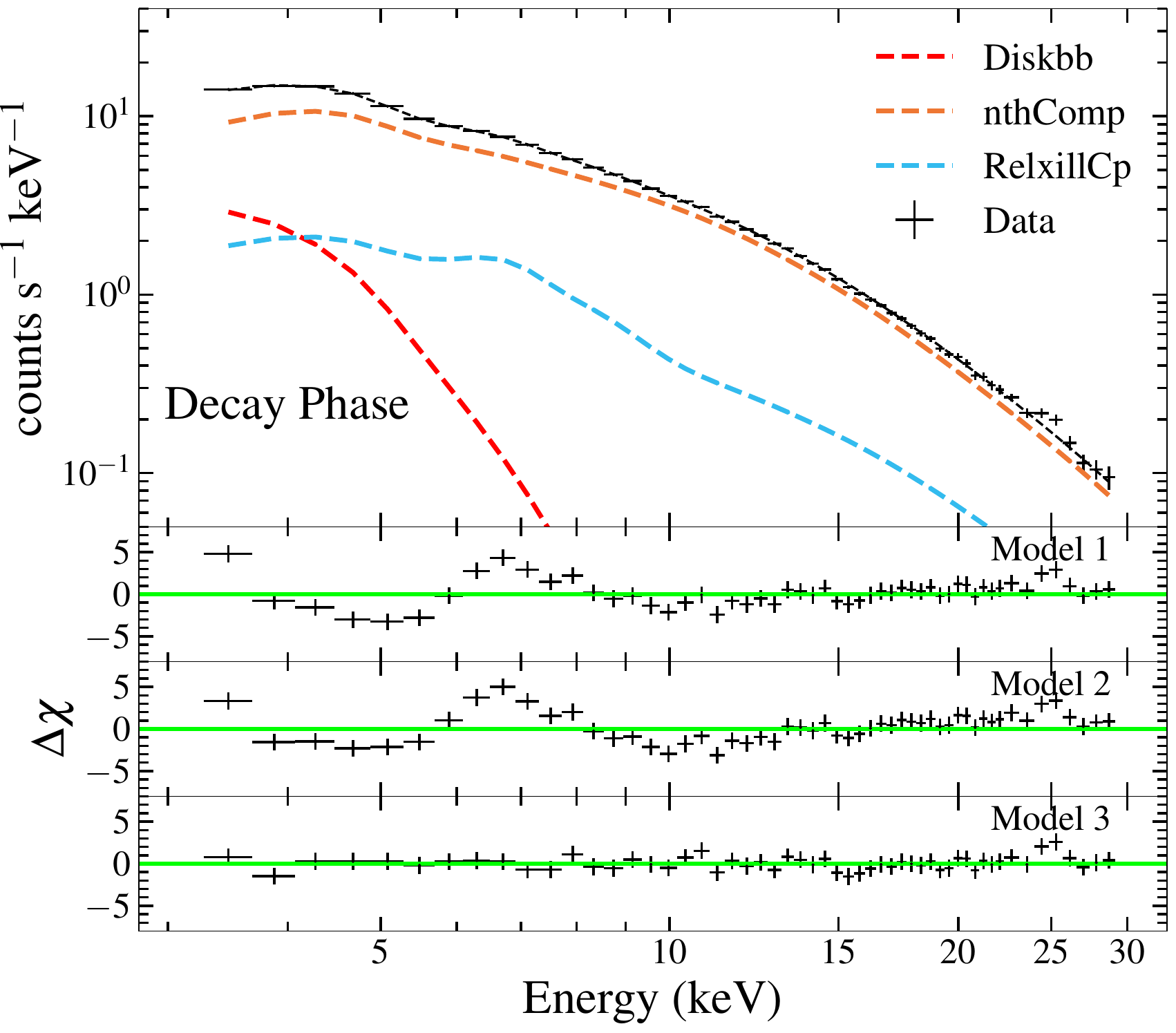}
\end{minipage}
    \caption{Representative \emph{RXTE}/PCA energy spectra (the top panels) and fitting residuals (the three bottom narrow panels are plotted for Model 1, 2 and 3, respectively) for both of the rise phase (the left panel, obs.ID: 80146-01-30-00) and the decay phase (the right panel, obs.ID: 93427-01-04-00). The $diskbb$, $nthcomp$ and $relxillcp$ model are plotted in red, orange and blue, respectively.}
    \label{fig:3}
\end{figure*}

In our spectral fittings, the Galactic absorption effect is always included in our models by implementing $tbabs$ model \citep{2000ApJ...542..914W}. 
The hydrogen column density ($N_{\rm H}$) is fixed to $2.3\times10^{22}\ {\rm cm^{-2}}$ following \citet{2006ApJ...646..394M}. 
Firstly, we start with a simple model consists of a power-law and a disc blackbody component (Model 1: $tbabs\times(diskbb+powerlaw)$) and obtain  large reduced-$\rm\chi^2$ ($\gg2$) in the most observations.
Then we replace $powerlaw$ with $nthcomp$, a physically motivated thermal Comptonization model which describes the spectrum of Compton upscattering photons \citep{1996MNRAS.283..193Z,1999MNRAS.309..561Z}. 
The spectral fitting with the second model (Model 2: $tbabs\times(diskbb+nthcomp)$) gives a smaller reduced-$\rm\chi^2$ but, as shown in Fig.~\ref{fig:3}a and b, the residuals still show obvious structures at energies around the iron line ($\sim6.4$ keV) and Compton hump ($\sim10$--30 keV), providing evidence of a relativistic reflection component \citep[][and references therein]{2013ApJ...768..146G}. Hence we add a reflection model $relxillcp$ \citep{2014MNRAS.444L.100D,2014ApJ...782...76G} in our spectral fittings (Model 3: $tbabs\times(diskbb+nthcomp+relxillcp)$).

For the spectral fitting with Model 3, we link the seed photon temperature ($kT_{\rm bb}$) of $nthcomp$ with the inner disk temperature ($kT_{\rm in}$) of $diskbb$ and choose the flavor of disk-blackbody seed photons for $nthcomp$. 
Since $relxillcp$ calculates the reflection component using the $nthcomp$ continuum, it is self-consistent to link the relevant parameters like the photon index ($\Gamma$) and electron temperature ($kT_{\rm e}$) between two models.
In the reflection model $relxillcp$, we assume the canonical power-law emissivity profile, $\epsilon\propto r^{-3}$, for the disc \citep{1989MNRAS.238..729F}. 
The parameter $R_{\rm out}$ is the outer radius of the accretion disc which turns out to be not sensitive to the overall fitting and hence is frozen at the maximum value ($1000 R_{\rm g}$, where $R_{\rm g}=GM/c^2$ is the gravitational radius). 
In order to make $relxillcp$ only calculate the reflection component, we fix the reflection fraction ($R_{\rm f}$) to $-1$. 
Referring to previous studies, the disc inclination ($i$), spin parameter of the black hole ($a_*$) and iron abundance ($A_{\rm Fe}$, in solar units) are set to $75^{\circ}$, 0.2 and 3.0, respectively \citep{2012ApJ...745L...7S,2020ApJ...893..142C}. The shape of $nthcomp$ continuum is set by the combination of the electron temperature ($kT_{\rm e}$) and scattering optical depth ($\tau_{\rm s}$), where the higher cut-off energy is parameterized by $kT_{\rm e}$ \citep{1999MNRAS.309..561Z}. However, in many observations, the cut-off energy is beyond the energy band of our spectral analysis (3--30 keV). 
Although $kT_{\rm e}$ can influence the reflection hump at energies around 20--40 keV which gives a chance to estimate it beyond the spectral coverage \citep[see][]{2015ApJ...808L..37G}, we note that $kT_{\rm e}$ is not completely reliable in our spectral fitting. For a less constrained $kT_{\rm e}$, we fix it to 300 keV.
Additionally, we notice the $diskbb$ contribution is marginal in where its parameters are not well constrained by using only PCA data and the spectral fitting does not even need this component in some observations \citep[see also][]{2014MNRAS.442.1767P,2019MNRAS.486.2705A}. For these observations, we fix the parameter $kT_{\rm in}$ and normalization of $diskbb$ at 0, then let $kT_{\rm bb}$ of $nthcomp$ as a free parameter in the fitting. 
We calculate the unabsorbed flux for each component in the energy range 3--30 keV by convolving $cflux$ model with these required models. 
After fitting energy spectra with Model 3, we obtain a reasonable reduced-$\chi^2$ ($\sim1$) for most observations (see Appendix~\ref{sec:A1}) and compute the 90 per cent confident-level uncertainties using the Markov Chain Monte Carlo (MCMC) technique, with length 40 000. The best-fit parameters of the two representative spectra presented in Fig.~\ref{fig:3} with Model 3 are summarized in Table~\ref{tab:1}.

\begin{table}
\caption{Best-fit Parameters and the Corresponding 90\% Confidence Intervals Obtained from Fitting the Representative \emph{RXTE}/PCA Spectra Presented in Fig.~\ref{fig:3} Using the Model 3: $tbabs\times(diskbb+nthcomp+relxillcp)$.}
\label{tab:1}
\begin{center}
\begin{tabular}{lcccr}
\hline \hline
Model & Parameters & Rise Phase & Decay Phase\\
$tbabs$ & $N_{\rm H}$ ($10^{22}\rm cm^{-2}$) & \multicolumn{2}{c}{2.3 (fixed)}\\
$relxillcp$ & $i$ (Deg) & \multicolumn{2}{c}{75 (fixed)}\\
& $a_*$ & \multicolumn{2}{c}{0.2 (fixed)}\\
& $R_{\rm out}\ (R_{\rm g})$ & \multicolumn{2}{c}{1000 (fixed)}\\
& $q^{\rm a}$ & \multicolumn{2}{c}{3 (fixed)}\\
& $A_{\rm Fe}$ (Solar Units) & \multicolumn{2}{c}{3 (fixed)}\\
& $R_{\rm f}$ & \multicolumn{2}{c}{$-1$ (fixed)}\\
\hline
$diskbb$ & $T_{\rm in}$ (keV) & $0.76^{+0.05}_{-0.05}$ & $0.62^{+0.08}_{-0.18}$\\
& $N_{\rm disk}^{\rm b}$ &  $866^{+280}_{-201}$ & $99^{+842}_{-53}$\\
$nthcomp$ & $\Gamma$ & $2.29^{+0.04}_{-0.03}$ & $1.83^{+0.03}_{-0.05}$\\
& $kT_{\rm e}$ (keV) &  $8.6^{+1.0}_{-0.5}$ & 300(fixed)\\
& $N_{\rm nth}^{\rm c}$ & $2.52^{+0.29}_{-0.21}$ & $0.15^{+0.05}_{-0.08}$\\
$relxillcp$ & $R_{\rm in}\ (R_{\rm ISCO})$ & $17.9^{+76.8}_{-8.1}$ & 100 (fixed)\\
& $\log_{10}(\xi)^{\rm d}$& \multirow{2}*{$2.81^{+0.13}_{-0.17}$} & \multirow{2}*{$3.53^{+0.67}_{-0.31}$}\\
& $(\rm erg\ cm\ s^{-1})$& &\\
& $N_{\rm rel}^{\rm e}\ (10^{-3})$ & $42.0^{+17.1}_{-9.7}$ & $2.3^{+4.50}_{-0.10}$\\
\hline
& $\chi^2/{\rm d.o.f}$ & 50.18/46 & 38.29/48 \\
\hline
\end{tabular}
\end{center}
\tablecomments{$^{\rm a}$ The power-law index of the emissivity profile ($\epsilon\propto r^{-q}$). $^{\rm b}$ Normalization of $diskbb$ model. $^{\rm c}$ Normalization of $nthcomp$ model. $^{\rm d}$ Log of the ionization parameter ($\xi$) of the accretion disc, where $\xi=L/nR^2$, with $L$ as the ionizing luminosity, $n$ as the gas density, and $R$ as the distance to the ionizing source. $^{\rm e}$ Normalization of $relxillcp$ model.}
    
\end{table}

\subsection{Timing-spectral Joint Analysis}
\label{sec:3.3}
The timing and spectral analyses presented in Section~\ref{sec:3.1} and~\ref{sec:3.2}, respectively,  provide the essential inputs for a spectral-timing joint diagnostic of the outburst what is shown in follows.

\subsubsection{Correlations between the QPO rms and non-thermal component}

\begin{figure*}
\centering
\begin{minipage}[c]{0.45\textwidth}
\centering
    \includegraphics[width=\linewidth]{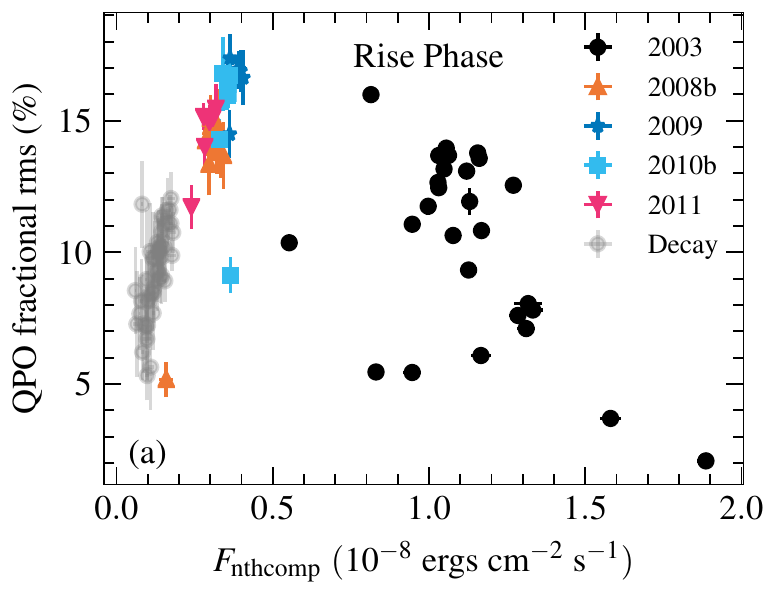}
\end{minipage}%
\begin{minipage}[c]{0.45\textwidth}
\centering
    \includegraphics[width=\linewidth]{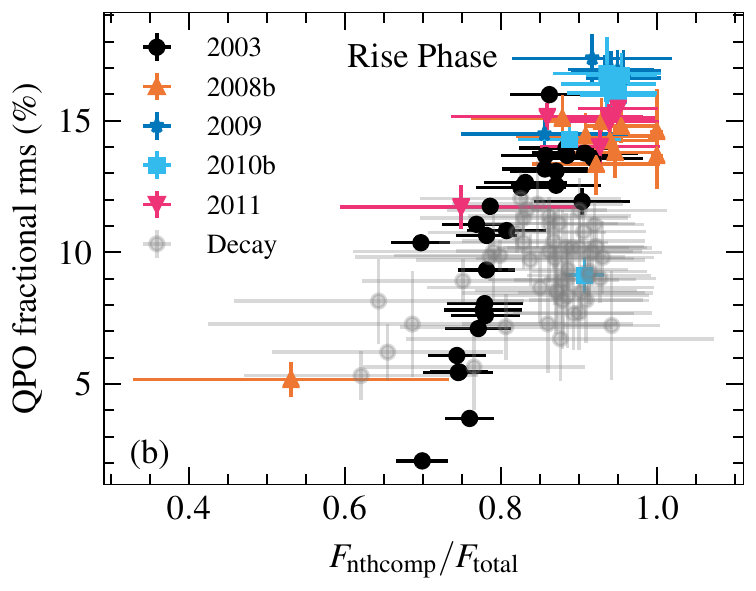}
\end{minipage}\\
\begin{minipage}[c]{0.45\textwidth}
\centering
    \includegraphics[width=\linewidth]{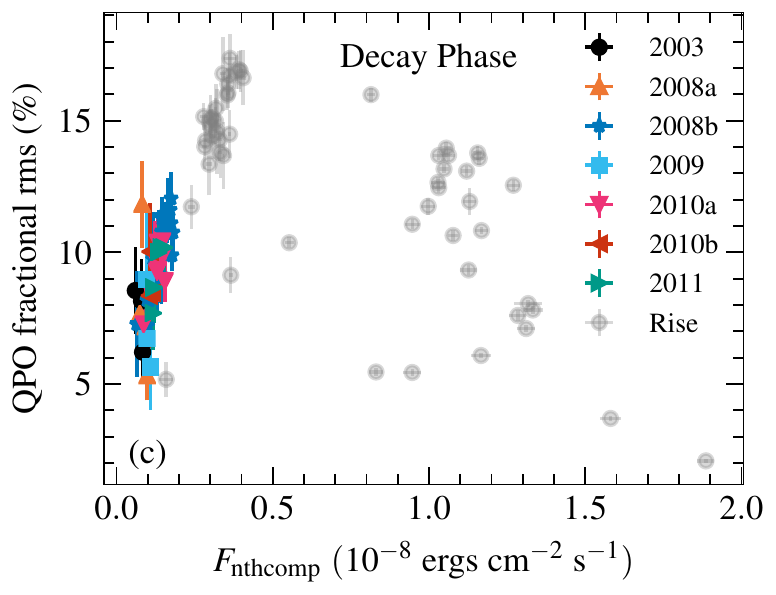}
\end{minipage}%
\begin{minipage}[c]{0.45\textwidth}
\centering
    \includegraphics[width=\linewidth]{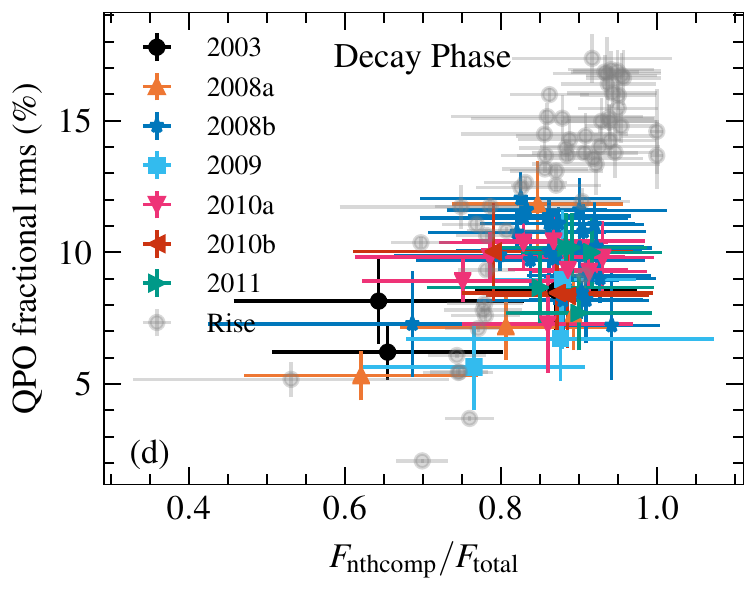}
\end{minipage}%
    \caption{The relations between QPO fractional rms and the non-thermal component. For clarity's sake, data of the different outburst rise phases are displayed in different colors in the panels (a) and (b), and data points of the decay phases are plot in gray without separating different outbursts, while the panels (c) and (d) display the data points of the rise and decay phases in the opposite way.}
    \label{fig:4}
\end{figure*}

Fig.~\ref{fig:4} presents the relations between the QPO fractional rms and non-thermal component during the rise and decay phases, separating the different outbursts. 
For clarity's sake, data points of the rise phase from different outbursts are displayed in different colors and shapes in Fig.~\ref{fig:4}a and b, and the data points of the decay phase are plotted in gray without separating different outbursts, while in the Fig.~\ref{fig:4}c and d, we display data sets in the opposite way. 
As shown in Fig.~\ref{fig:4}a, in the rise phase, there are no remarkable correlations between the fractional rms and non-thermal fluxes ($F_{\rm nthcomp}$), and data points from different outbursts distribute widely in $F_{\rm nthcomp}$ which is consistent with the large differences in the outburst peak fluxes shown in Fig.~\ref{fig:1}. 
However, if we display the QPO fractional rms as a function of the non-thermal fraction ($F_{\rm nthcomp}/F_{\rm total}$), the positive correlations between the fractional rms and non-thermal component become significant and consistent among different outbursts (see Fig.~\ref{fig:4}b). 
In the decay phase, different outbursts show the similar non-thermal fluxes ($\rm\sim 10^{-9}\ ergs\ cm^{-2}\ s^{-1}$). 
However, compared with the rise phase, the positive correlation between the fractional rms and non-thermal fraction is not clear within the decay phase because of the relatively larger error bar of the non-thermal fraction (see Fig.~\ref{fig:4}c and d).    

\subsubsection{QPO rms dependence on frequency}

\begin{figure*}
\centering
\begin{minipage}[c]{0.45\textwidth}
\centering
    \includegraphics[width=\linewidth]{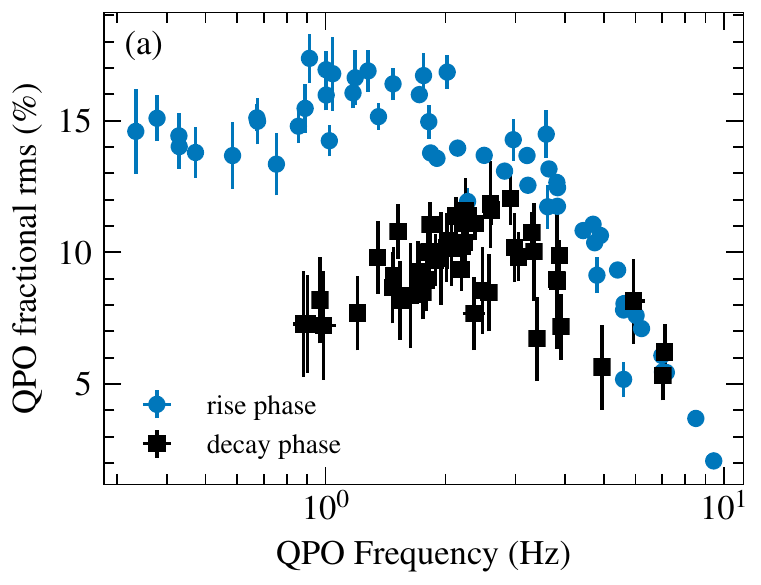}
\end{minipage}%
\begin{minipage}[c]{0.45\textwidth}
\centering
    \includegraphics[width=\linewidth]{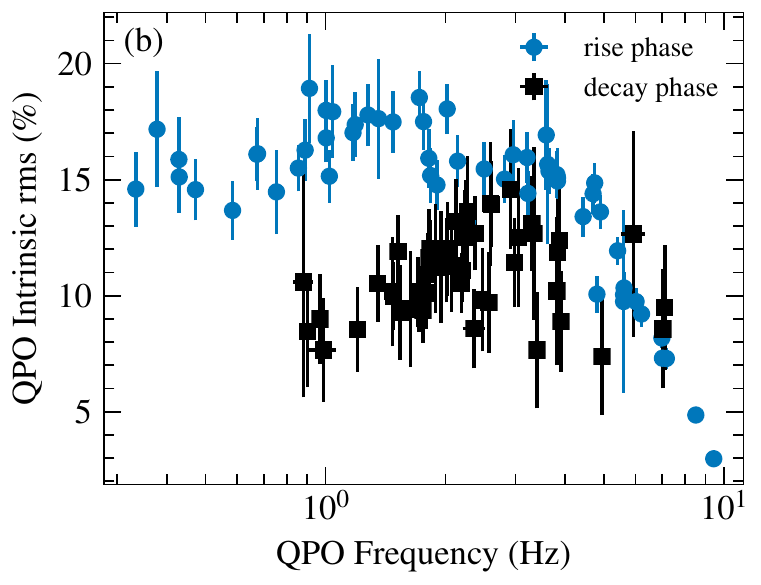}
\end{minipage}
    \caption{QPO fractional rms (the left panel) and intrinsic rms (the right panel) plotted as a function of frequency. The rise and decay phase are distinguished in different colors and shapes: data points from the rise phase are plotted as blue circles, while those from the decay phase are plotted as black squares.}
    \label{fig:5}
\end{figure*}

Type C QPOs usually appear in the HIMS, a stage shows the significant transition in both of the timing and spectral domains \citep{2005Ap&SS.300..107H,2006ARA&A..44...49R}, where properties of the type C QPO, like central frequency and fractional rms, evolve in a large value range. 
Previous studies have presented that the fractional rms of type C QPOs varies with the frequency \citep[see][]{2015MNRAS.447.2059M,2020MNRAS.496.5262V,2020MNRAS.494.1375Z,2022MNRAS.512.4541W}. 
Here, we analyse a number of type C QPO samples across different outbursts of H 1743--322 and present the QPO fractional rms dependence on fundamental frequency in both the rise and decay phases in Fig.~\ref{fig:5}a for comparisons.
Data points displayed are not be distinguished among different outbursts, but only between the rise and decay phases.
As one can see, during the rise phase, the QPO fractional rms increases slightly with frequency below 1 Hz and then remains roughly flat around 1--2 Hz, while shows a significant drop at higher frequencies.
However, the presented relation of the decay phase deviates from that of the rise phase at low frequencies, where the deviation becomes larger for the lower frequency. 
Also, these relations between the QPO fractional rms and frequency in both of the rise and decay phases are consistent among different outbursts. The fractional rms presented in the study is computed by adding in quadrature the rms of the harmonic peaks. However, we have checked the case that taking the rms of QPO fundamental only, and find the dependence of fractional rms on frequency deviates only slightly with respect to the results presented in Fig.~\ref{fig:5}, where the two outburst stages remain different branches in the relations between QPO fractional rms and frequency clearly.

Since the fractional rms is defined as the fractional variability of the flux, the fractional rms we compute in the Section~\ref{sec:3.1} and presented in Fig.~\ref{fig:5}a can be described by
\begin{equation}
{\rm rms}=\frac{F_{\rm var}}{F_{\rm total}}=\sigma\times\frac{F_{\rm c}}{F_{\rm total}},
\end{equation}
where $F_{\rm var}$ is the variable flux (absolute rms), $F_{\rm total}$ is the total time-averaged flux, $\sigma$ is a function for the intrinsic rms, and $F_{\rm c}$ represents the flux contributing to the QPO variability \citep[see also][]{2020JHEAp..25...29K,2021MNRAS.508..287S}. If we assume that only the non-thermal component contributes to the type C QPO, while the fractional rms is diluted by the other components, i.e. the thermal and reflected components, then $F_{\rm c}$ is equal to $F_{\rm nthcomp}$. So the intrinsic rms is the variability amplitude of non-thermal emission, which can be computed by
\begin{equation}
\label{eq:3}
\sigma={\rm rms}\times\frac{F_{\rm total}}{F_{\rm nthcomp}}.
\end{equation}
The dependence of the intrinsic rms on frequency is presented in Fig.~\ref{fig:5}b, and we find it is similar to that of the fractional rms displayed in Fig.~\ref{fig:5}a. It is clear to see the two separate branches of the rise and decay phases.

\subsection{Coronal Parameters and Radio Emission}

\begin{figure*}
\centering
\begin{minipage}[c]{0.45\textwidth}
\centering
    \includegraphics[width=\linewidth]{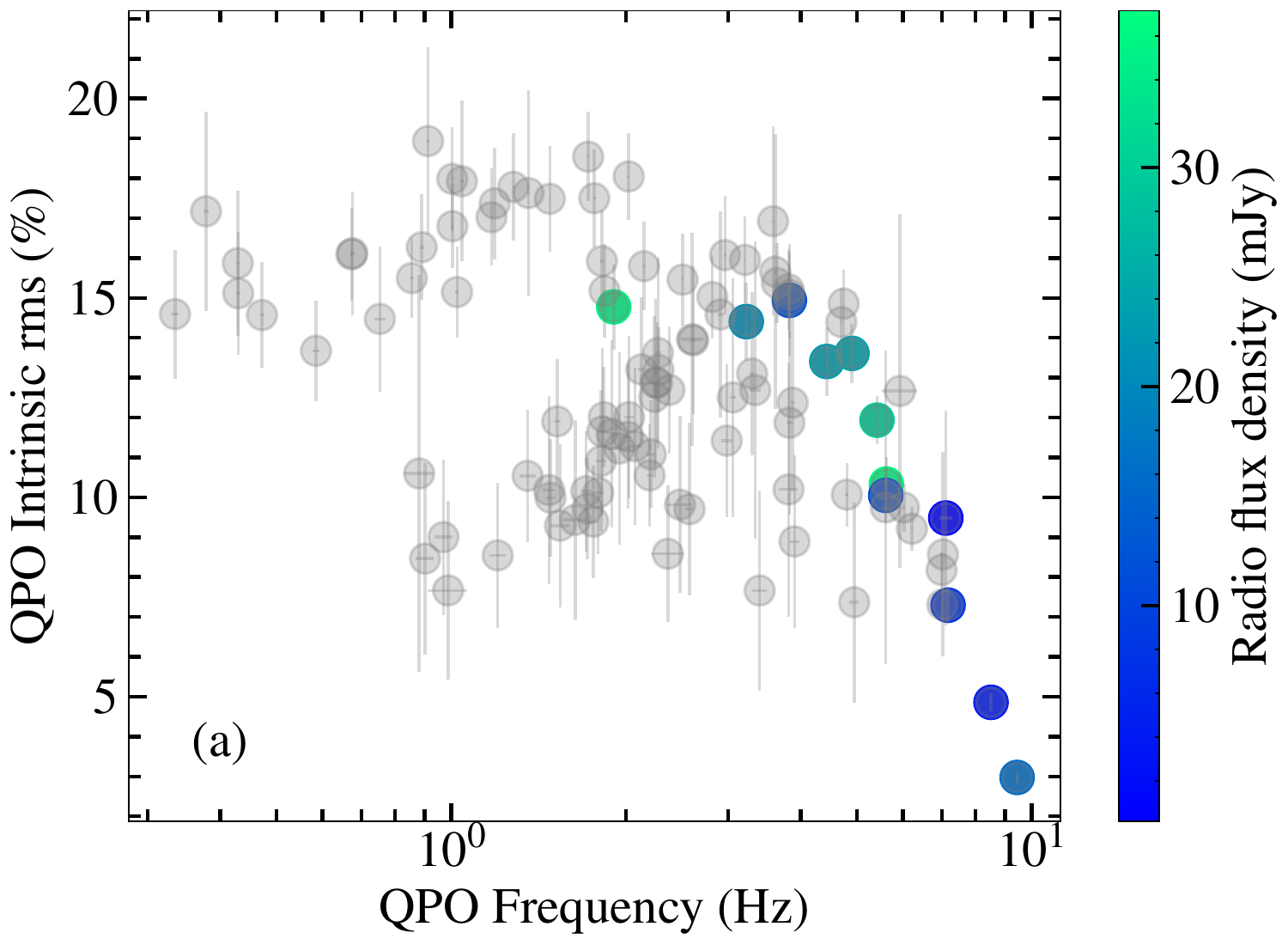}
\end{minipage}%
\begin{minipage}[c]{0.45\textwidth}
\centering
    \includegraphics[width=\linewidth]{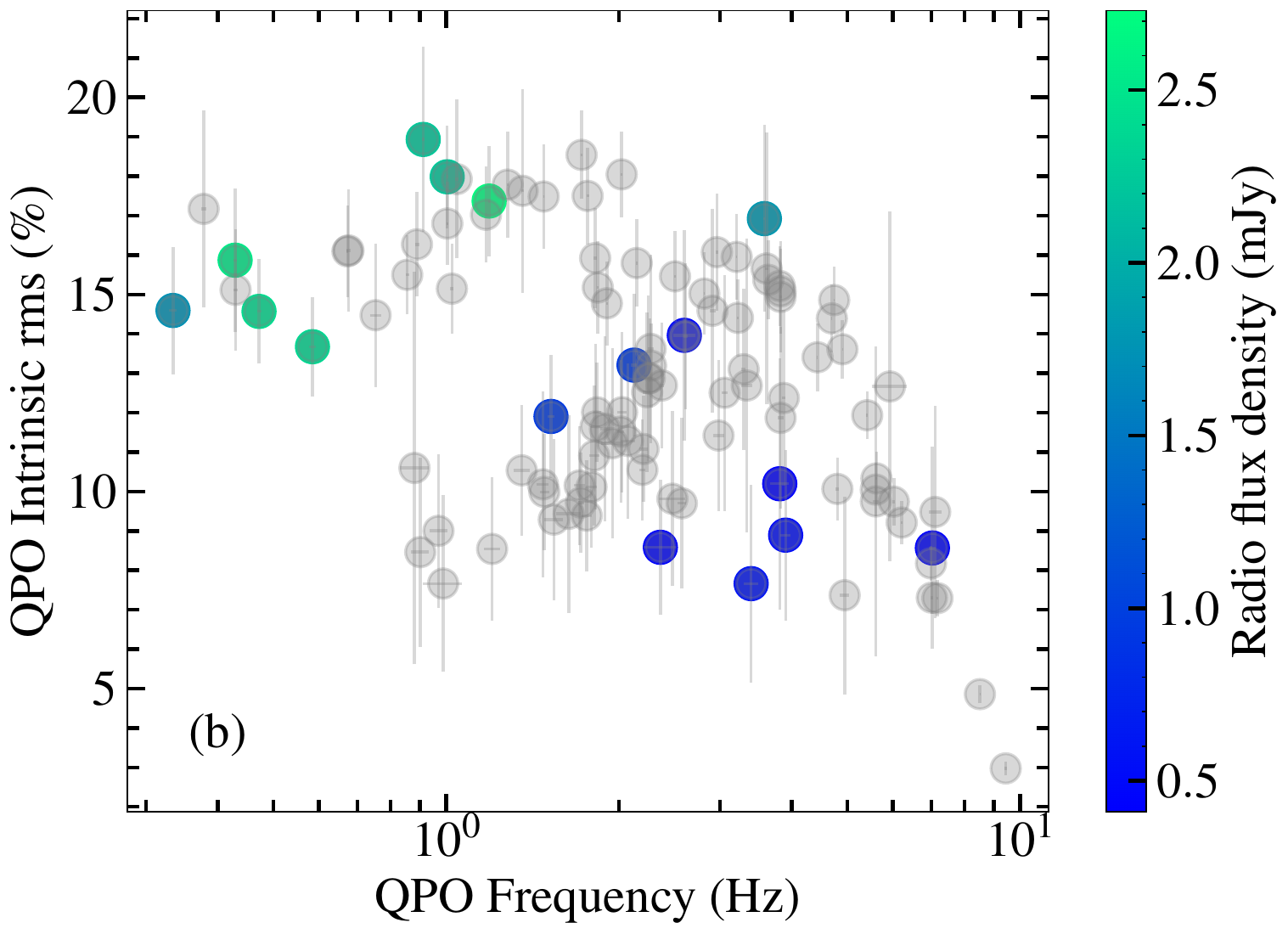}
\end{minipage}\\
\begin{minipage}[c]{0.45\textwidth}
\centering
    \includegraphics[width=\linewidth]{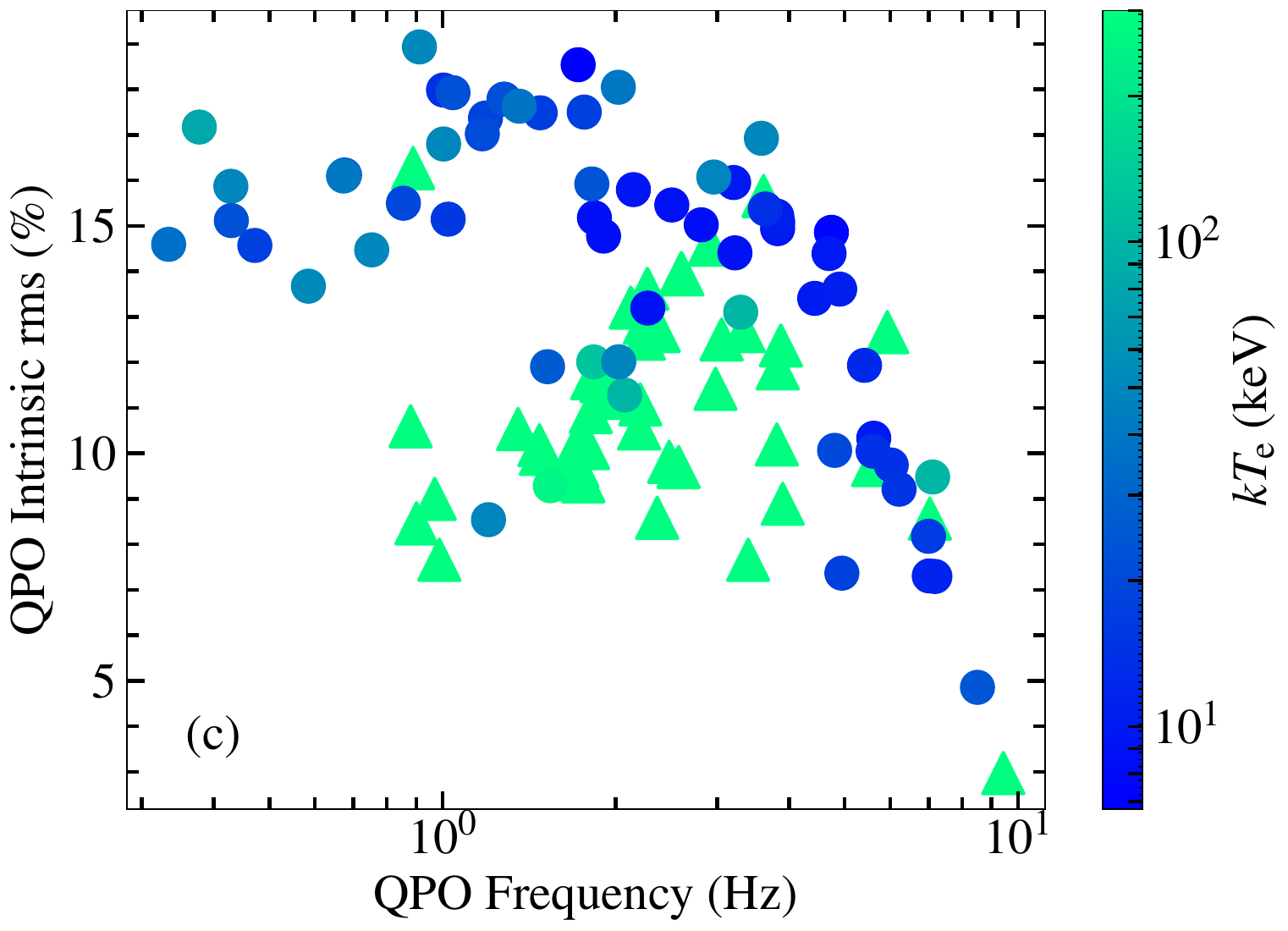}
\end{minipage}%
\begin{minipage}[c]{0.45\textwidth}
\centering
    \includegraphics[width=\linewidth]{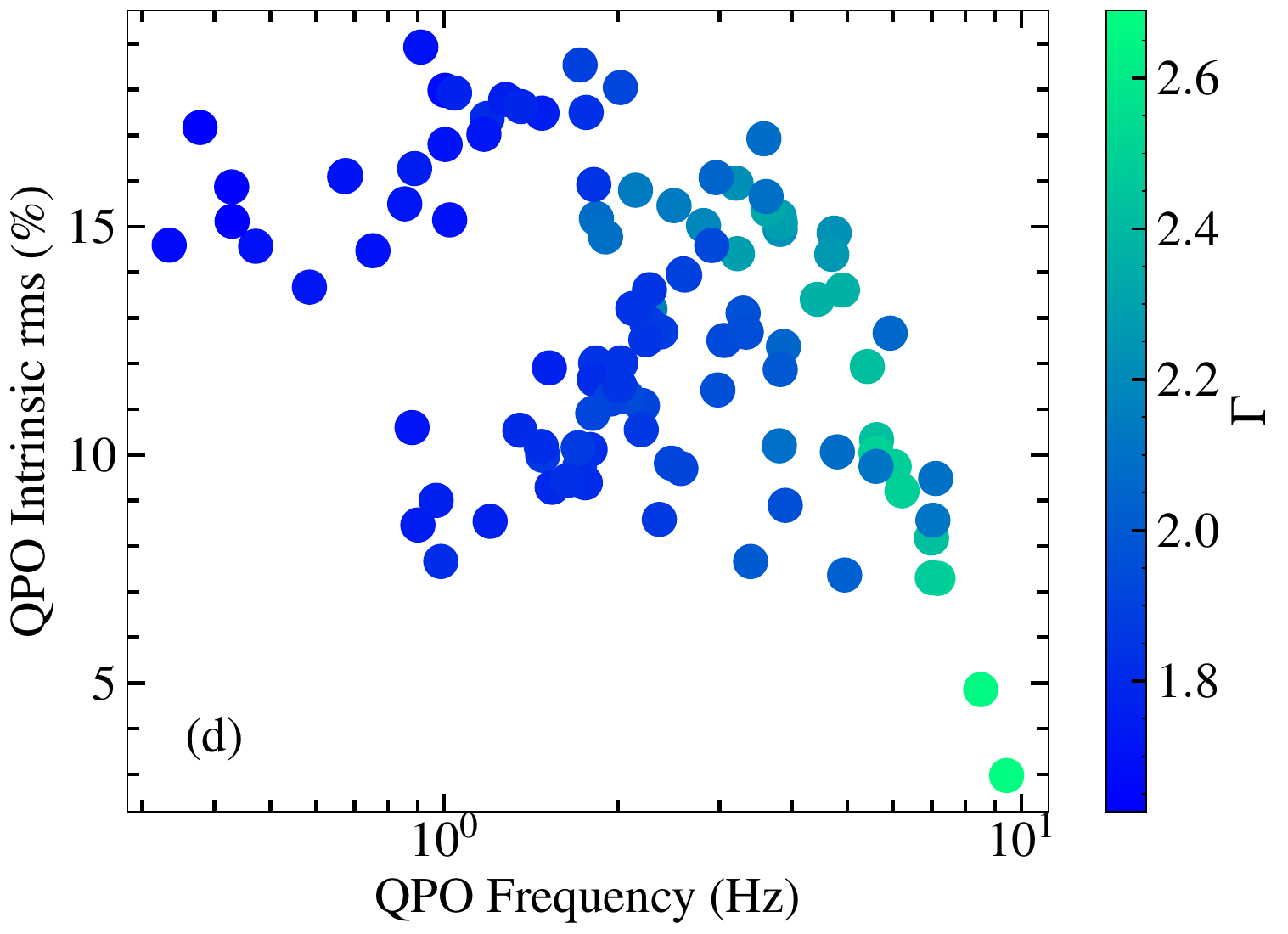}
\end{minipage}%
    \caption{The QPO intrinsic rms plotted as a function of frequency. The shade of the data points indicate the quasi-simultaneous measurements of the radio flux density (a and b), $kT_{\rm e}$ (c) and $\Gamma$ (d), respectively. In panels (a) and (b), we plot the entire QPO data set from our timing analysis in gray, only these data points with radio observations are colored. Furthermore, since the outburst 2003 is much brighter than the other outbursts, we present the radio flux density of the 2003 outburst in panel (a) alone, while those of the other outbursts are shown in panel (b). In panel (c), these triangles indicate that $kT_{\rm e}$ is too high to be constrained using \emph{RXTE}/PCA energy spectra, hence fixed at 300 keV.}
    \label{fig:6_p}
\end{figure*}

We have shown that the dependence of the QPO intrinsic rms (hereafter QPO rms) on frequency shows two branches: the QPO rms in the outburst rise phase is significantly larger than that in the decay phase at low frequencies. To investigate this bi-modality in more detail, we present the dependence of the rms-$f_{\rm QPO}$ relation on radio fluxes and spectral parameters in Fig.~\ref{fig:6_p}. The quasi-simultaneous measurements of the radio flux density at $\sim8.5$ GHz ($S_{\nu=8.5{\rm GHz}}$) are taken from \citet{2009ApJ...698.1398M,2010MNRAS.401.1255J,2011MNRAS.414..677C,2012MNRAS.421..468M}. For details of the radio observations used in the present study, see Appendix \ref{sec:A4}. We note that not all \emph{RXTE}/PCA observations in this study have quasi-simultaneous radio measurements. However, for clearly showing the radio dependence of the two different branches, we plot the entire QPO data set from our timing analysis in gray in the top two panels of Fig.~\ref{fig:6_p}, while coloring these data points which have radio observations. Furthermore, since 2003 outburst is much brighter than the other outbursts, we present the radio data of 2003 outburst in Fig.~\ref{fig:6_p}a alone, while those of the other outbursts are shown in Fig.~\ref{fig:6_p}b. The dependence of the coronal temperature ($kT_{\rm e}$) and photon index ($\Gamma$) are presented in the bottom two panels of Fig.~\ref{fig:6_p}, respectively. The shade of the data points in each panel indicate the measured values of the parameters. Radio emission during the outburst rise phase of 2003 outburst is the strongest ($S_{\nu=8.5{\rm GHz}}>10\ {\rm mJy}$), while that of the other outbursts is relatively weaker ($S_{\nu=8.5{\rm GHz}}\sim2.5\ {\rm mJy}$). For an individual outburst, there is a marginally decreasing trend of the radio flux density from low QPO frequency ($f_{\rm QPO}<2$ Hz) to high frequency ($f_{\rm QPO}\sim8$ Hz). Additionally, radio emission of the decay phase ($S_{\nu=8.5{\rm GHz}}<1\ {\rm mJy}$) is significantly weaker than that of the rise branch. In contrast to the radio flux, the electron temperature is higher in the decay branch. These triangles plotted in Fig.~\ref{fig:6_p}c indicate that $kT_{\rm e}$ are too high to be constrained using \emph{RXTE}/PCA energy spectra, hence fixed at 300 keV (see Section~\ref{sec:3.2} for details). However, for the photon index, there are no apparent differences between the two branches in the similar frequency range.

\subsection{Phenomenological Analysis with the L-T Precession Model}
\label{sec:3.4}
The phenomenon that the QPO rms is dependent on the frequency and outburst stage (see Fig.~\ref{fig:5}) indicates that the intrinsic properties of QPOs changed during the outburst.
If type C QPOs are produced by the L-T precession of the entire corona \citep[see][]{2009MNRAS.397L.101I,2018ApJ...858...82Y}, a simplified geometrical model presented as follows can be used to estimate the intrinsic rms semi-quantitatively. 

\begin{figure}
\centering
\includegraphics[width=\linewidth]{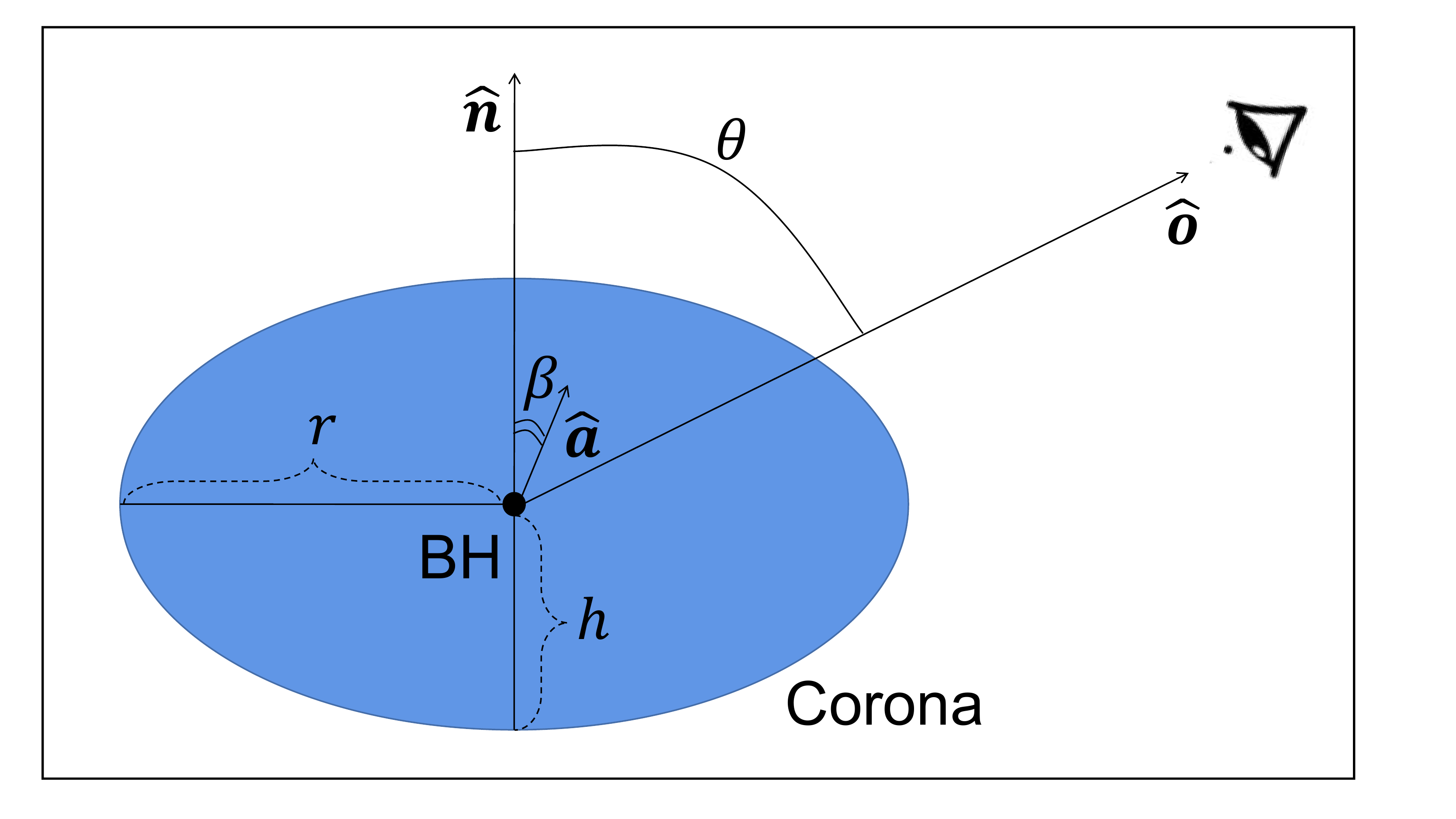}
\caption{Schematic diagram illustrating the cross-section of the corona. $\hat{\bm n}$, $\hat{\bm o}$ and $\hat{\bm a}$ represent the normal of the corona, unit vector pointing from the black hole to the observer and unit vector of black hole spin. respectively. $\beta$ is the angle between binary orbit and black hole spin, and $\theta$ is the included angle between $\hat{n}$ and $\hat{o}$. The coronal shape is described by the height scale $h/r$, the ratio of the semiminor and semimajor axis of the ellipse. We note this presented representative geometry is a case of $\Phi=0$.}
\label{fig:7}
\end{figure}

Using the coordinate system described by \cite{2013ApJ...778..165V,2015ApJ...807...53I}, the unit vector pointing from the central black hole to the observer is given by
\begin{equation}
\hat{\bm{o}}=\left(\sin{i}\cos{\Phi},\sin{i}\sin{\Phi},\cos{i}\right),
\end{equation}
where $i$ is the binary inclination and $\Phi$ is the azimuth of the observer. 
The instantaneous normal of the corona is $\hat{\bm{n}}$, changing with the precession phase angle $\omega$:
\begin{align}
\hat{\bm{n}}=(&\sin{\beta}\cos{\beta}\left(1+\cos{\omega}\right),\notag\\
&\sin{\beta}\sin{\omega},\cos^2{\beta}-\sin^2{\beta}\cos{\omega}),
\end{align}
where $\beta$ is the angle between binary orbit and black hole spin. We note that $\omega$ changes from 0 to $2\pi$ on the precession period, hence leads to the changes of $\hat{\bm{n}}$. 
Then cosine of the included angle ($\theta$) between $\hat{\bm{n}}$ and $\hat{\bm{o}}$ can be written by
\begin{align}
\cos{\theta}=&\hat{\bm{n}}\cdot\hat{\bm{o}}\notag\\
=&\sin{\beta}\cos{\beta}\sin{i}\cos{\Phi}\left(1+\cos{\omega}\right)\notag\\
&+\sin{\beta}\sin{\omega}\sin{i}\sin{\Phi}\notag\\
&+\cos{i}\left(\cos^2{\beta}-\sin^2{\beta}\cos{\omega}\right).
\end{align}
We consider a simplified problem with assuming that the coronal shape is a crushed sphere, i.e. viewed from the coronal normal ($\hat{\bm{n}}$) is a circle while the cross-section is an ellipse (see Fig.~\ref{fig:7}). 
The shape of the corona can be described by defining the scale height, $h/r$, where $h$ is the semiminor axis and $r$ is the semimajor axis of the ellipse, respectively. 
In the geometry described above, the projected area of the corona to the observer is
\begin{align}
S_{\rm ob}&=\pi r^2\left[\sin^2{\theta}\cdot\left(h/r\right)^2+\cos^2{\theta}\right]^{1/2}\notag\\
&=\pi r^2\left\{\left(h/r\right)^2+\left[1-(h/r)^2\right]\cos^2{\theta}\right\}^{1/2}.
\end{align}
For simplicity, we assume the radiation of the corona is homogeneous and isotropic, then the observed flux from the corona is a function of $\theta$ and $h/r$:
\begin{align}
\label{eq:8}
F&=\bar{I}\cdot\Delta\Omega=\bar{I}\cdot S_{\rm ob}/D^2\notag\\
&\propto\left\{\left(h/r\right)^2+\left[1-(h/r)^2\right]\cos^2{\theta}\right\}^{1/2}\left(1-e^{-\tau}\right),
\end{align}
where $\bar{I}$ is the average radiation intensity, $\Delta\Omega$ is the solid angle of the corona in the view of the observer, $D$ is the distance from the source to the observer, and $\tau$ is the effective optical depth, also a function of $\theta$ and $h/r$. 
We estimate $\tau$ as a simple case:
\begin{equation}
\label{eq:9}
\tau=\tau_0\left\{\left(h/r\right)^2+\left[1-\left(h/r\right)^2\right]\cos^2{\theta}\right\}^{-1/2},
\end{equation}
where $\tau_0$ is the minimum optical depth of the corona, i.e. viewed from the coronal normal (see Appendix~\ref{sec:A2} for details). Combining Equations \ref{eq:8} and \ref{eq:9}, we find that the flux from the corona changes with $\theta$ on the precession period. On the basis of the above, the minimum and maximum coronal flux could be calculated in the $\omega$ value range $0-2\pi$. Then the QPO rms can be estimated by
\begin{equation}
{\rm rms}=\frac{F_{\rm max}-F_{\rm min}}{F_{\rm max}+F_{\rm min}},
\end{equation}
where $F_{\rm max}$ and $F_{\rm min}$ are the maximum and minimum value of the coronal flux, respectively, on the precession period. 

\begin{figure*}
\centering
	\includegraphics[width=0.9\textwidth]{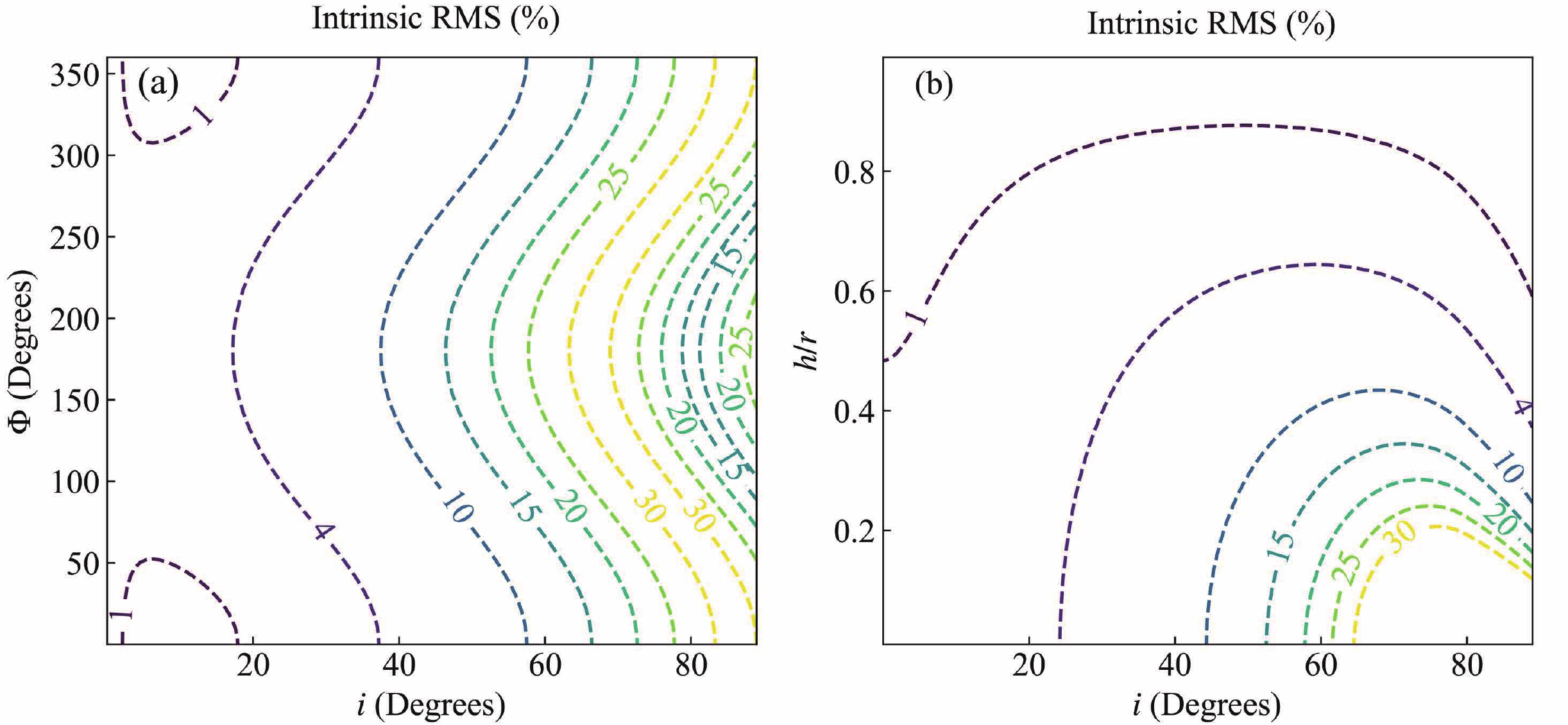}
    \caption{The calculated results using the geometrical model. In the calculation, we assume a translucent corona with $\tau_0=1$ which is consistent with the previous hypotheses and simulations \citep[see][]{2018ApJ...858...82Y,2019NewAR..8501524I,2020ApJ...897...27Y}. Panel (a) displays the intrinsic rms dependence on the entire range of viewing angles with a fixed $h/r$ value (0.2), panel (b) displays the dependence of intrinsic rms on $h/r$ with different inclinations ($\Phi$ is set to $90^\circ$).}
    \label{fig:8}
\end{figure*}

The calculated results using the simplified geometrical model are presented in Fig.~\ref{fig:8}. 
In our calculation, we assume a translucent corona with $\tau_0=1$ which is consistent with the previous hypotheses and simulations \citep[see][]{2015ApJ...807...53I,2018ApJ...858...82Y,2020ApJ...897...27Y}. 
Fig.~\ref{fig:8}a displays the intrinsic rms dependence on the entire range of viewing angles ($h/r=0.2$), where it shows that rms increases with the inclination ($i$), while gets the peak values at $\Phi=180^{\circ}$. 
However, $i$ and $\Phi$ are constant for a specific black hole source, so the scale height ($h/r$) is the only parameter to influence intrinsic rms during the outburst evolution.
In Fig.~\ref{fig:8}b, we find that QPO rms is highly dependent on $h/r$, especially in the high inclination sources ($i\sim60^{\circ}$--$90^{\circ}$). We note that the general relativity effects, e.g. light bending, are not considered in the simplified L-T precession model, but the results presented in Fig.~\ref{fig:8} are consistent with previous studies which took the general relativity into account \citep[see][]{2013ApJ...778..165V,2015ApJ...807...53I,2018ApJ...858...82Y}. This fact indicates that GR effects, e.g. light bending, could be the second-order effects to affect the QPO rms in the L-T precession model.

\begin{figure*}
\centering
\begin{minipage}[c]{0.45\textwidth}
\centering
    \includegraphics[width=\linewidth]{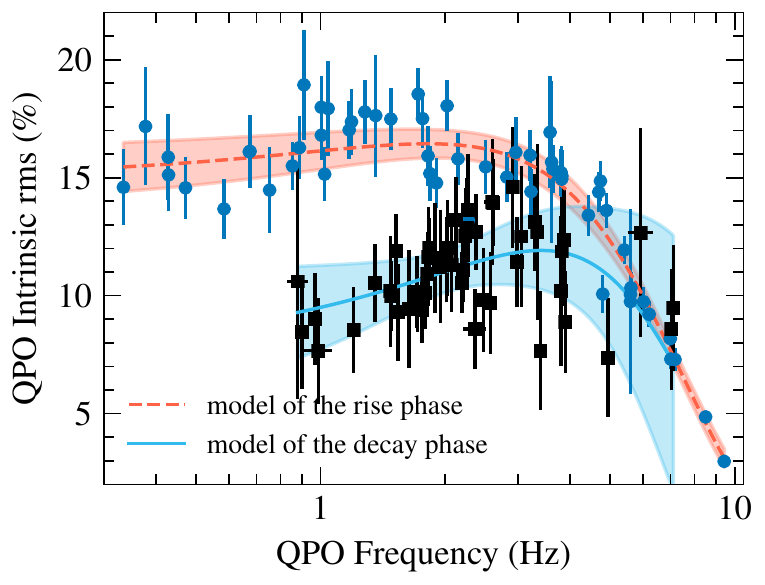}
\end{minipage}%
\begin{minipage}[c]{0.45\textwidth}
\centering
    \includegraphics[width=\linewidth]{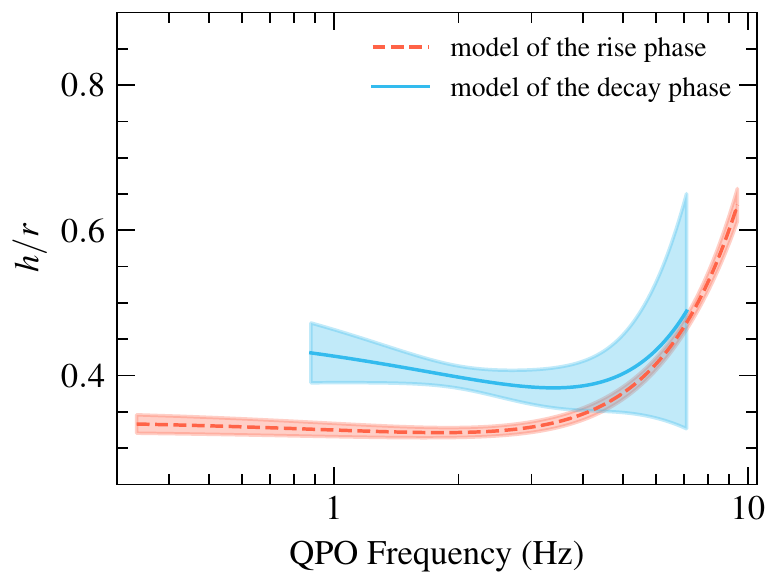}
\end{minipage}
    \caption{Fitting results of the relations between QPO intrinsic rms and frequency that presented in Fig.~\ref{fig:5}b with the simplified L-T precession model, where the trend lines and colored shaded regions represent the median and 3$\sigma$ confidence intervals of the fittings: the rise phase is displayed in red and decay phase is display in blue. The data points plotted in the left panel are same as that presented in Fig.~\ref{fig:5}b.}
    \label{fig:9}
\end{figure*}

Calculating with different values of $h/r$ can obtain different intrinsic rms, which allows us to fit the relations between QPO intrinsic rms and frequency presented in Fig.~\ref{fig:5}b with the model. From a phenomenological motivation, we assume $h/r$ as a quadratic function of $f_{\rm QPO}$:
\begin{equation}
\label{eq:13}
h/r=k_1\cdot f_{\rm QPO}^2 + k_2\cdot f_{\rm QPO} + k_3,
\end{equation}
for both of the rise and decay phases, where $f_{\rm QPO}$ is the QPO frequency, $k_1$, $k_2$ and $k_3$ are the model parameters. 
In this fitting, we fix the inclination, $i$, to $75^{\circ}$ which is consistent with our spectral analysis and the misalignment angle, $\beta$, to $10^{\circ}$ following \citet{2013ApJ...778..165V} and \citet{2015ApJ...807...53I}. 
Since the azimuth of observer, $\Phi$, is constant for a specific BH source, it only affects our estimation for the value of $h/r$, not the evolution trend, we fix it to a median ($90^{\circ}$). 
The MCMC technique is used to carry out the parameter estimation with the uniform prior distribution. 
The posterior probability distributions of the model parameters are presented in Appendix~\ref{sec:A3}. 
From the fitting, we get $k_1=(5.4\pm0.3)\times10^{-3}$, $k_2=(-1.96\pm0.27)\times10^{-2}$, $k_3=(3.39\pm0.05)\times10^{-1}$ and $\chi^2/{\rm d.o.f}=65.90/53$ for the rise phase, while $k_1=(7.5\pm2.9)\times10^{-3}$, $k_2=(-5.0\pm1.9)\times10^{-2}$, $k_3=(4.70\pm0.27)\times10^{-1}$ and $\chi^2/{\rm d.o.f}=28.02/47$ for the decay phase. 
The fitting results are displayed in Fig.~\ref{fig:9}, where the trend lines and colored shaded regions represent the median and 3$\sigma$ confidence intervals of the fitting, and the data points shown in the left panel are the same as those presented in Fig.~\ref{fig:5}b. 
In Fig.~\ref{fig:9}b, we display the $h/r$ dependence on $f_{\rm QPO}$ (described by Eq.~\ref{eq:13}). The evolution trends of $h/r$ are consistent between the two phases at high frequencies ($f_{\rm QPO}>4$ Hz), which show the increase with the increasing frequency.
At low frequencies, $h/r$ of both the two phases are negatively correlated to the frequency, but $h/r$ values and the gradient of the decay phase are significantly larger than those of the rise phase.

\section{Discussion} \label{sec:4}
We have systematically investigated the properties of type C QPOs by analysing one hundred and six \emph{RXTE}/PCA observations across multiple outbursts of H 1743--322 from 2003 to 2011. Fig.~\ref{fig:4} shows large variances of non-thermal fluxes among different outbursts, which distort the correlation between the non-thermal flux and the QPO fractional rms.
However, the QPO fractional rms of different outbursts is positively correlated to the non-thermal fraction ($F_{\rm nthcomp}/F_{\rm total}$) consistently. Additionally, the co-evolution between the QPO intrinsic rms and frequency keeps similar traces among different outbursts (see Fig.~\ref{fig:5}).
Since different outbursts could have very different luminosity levels, these consistent behaviours across outbursts indicate that the QPO intrinsic rms (hereafter QPO rms) is independent on the individual outburst brightness. However, the dependence of the QPO rms on frequency can be classified into two branches, where QPO rms in the outburst rise phase is significantly higher than that in the decay phase at low frequencies. Radio observations and X-ray spectral analyses reveal more differences between the two branches.

\subsection{Trace the Coronal Geometry with the Simplified L-T Precession Model}
This phenomenon that the QPO rms is independent on the outburst brightness has been also reported in other sources \citep[e.g. GX 339--4,][]{2021MNRAS.508..287S}. If type C QPOs are produced by L-T precession of the corona, the X-ray QPO variability owes to the geometric wobble of the corona, which changes the projection area of the corona with respect to the observer to modulate the X-ray flux \citep{2009MNRAS.397L.101I,2015ApJ...807...53I,2018ApJ...858...82Y,2020ApJ...897...27Y}. 
Calculations presented in Section~\ref{sec:3.4} give that the intrinsic amplitude is highly dependent on the coronal shape, $h/r$, for a specific BH source (assuming a translucent corona, $\tau\sim1$).
Accordingly, the similar QPO rms amplitudes with different outburst intensities indicate the coronal shape may not depend on the individual outburst brightness. 

The left panel of Fig.~\ref{fig:9} shows that, in the rise phase, the QPO rms increases slightly at low frequencies, while decreases sharply after reaching the peak value at $\sim2$ Hz. In the dynamic part of the L-T precession model, $f_{\rm QPO}$ is negatively correlated to the outer radius, $r$, of the corona \citep{2009MNRAS.397L.101I}, so the increasing $f_{\rm QPO}$ indicates a decreasing $r$. A possible explanation for the dependence of the QPO rms on frequency is that when the coronal outer radius ($r$) evolves, the coronal shape, i.e. $h/r$, changes synchronously. The fitting results of the rms-$f_{\rm QPO}$ relation with the simplified L-T precession model show that $h/r$ of the rise and decay phases are consistent at high frequencies ($f_{\rm QPO}>3$ Hz), while $h/r$ of the decay phase is larger than that of the rise phase in the low frequency range (see Fig.~\ref{fig:9}). The spectral analysis shows that the coronal temperature ($kT_{\rm e}$) of the decay phase is obviously higher than that of the rise phase (see Fig.~\ref{fig:6_p}c). If the corona is a hot accretion flow (e.g. Advection-Dominated Accretion Flow, ADAF), it is mainly supported by the gas pressure \citep[][]{1994ApJ...428L..13N,2014ARA&A..52..529Y,2022arXiv220106198L}, and the height ($h$) could hence be lager with the higher coronal temperature. Accordingly, the corona in the decay phase could have relatively larger $h/r$ at a specific frequency, and then precesses with a lower variability amplitude. This is because the variability of the coronal projection area with respect to the observer is smaller with a higher $h/r$ value (see Fig.~\ref{fig:8}).

\subsection{Qualitative Interpretation with the Time-dependent Comptonization Model}
Time-dependent Comptonization models can explain quantitatively the rms spectrum and the phase lag spectrum of QPOs, by requiring coupled oscillations of the physical quantities: coronal temperature, $kT_{\rm e}$, temperature of the source of seed photons, $kT_{\rm s}$ and the external heating rate, $\dot{H}_{\rm ext}$ \citep[see][]{2020MNRAS.492.1399K,2022MNRAS.515.2099B}. Although the QPO dynamic origin is not specified in these models, the recent proposal of \citet[][]{2022A&A...662A.118M} that the QPO frequency arises from a resonance between the hot Comptonizing corona and the colder accretion disc via the coupling of the energy gains and losses in the system, could provide the dynamic part to the time-dependent Comptonization models, since the disc-corona coupling is also the mechanism suggested by these models to explain the QPO radiative properties. In the time-dependent Comptonization models, the fractional variability amplitude of QPOs (QPO fractional rms) is normalized by the variability amplitude of the external hating rate, $\delta \dot{H}_{\rm ext}$, which is a fitting parameter. The energy-averaged rms in a specific energy range is therefore dependent on both the shape and the normalization ($\delta \dot{H}_{\rm ext}$) of the rms spectrum. \citet{2021MNRAS.503.5522K} and \citet{2022MNRAS.513.4196G} fitted the rms spectra and phase lag spectra of type C QPO observations of GRS 1915+105 and found $\delta \dot{H}_{\rm ext}$ is dependent on QPO frequency, where $\delta \dot{H}_{\rm ext}$ is positively correlated strongly to $f_{\rm QPO}$ at low frequencies ($f_{\rm QPO}<1$ Hz), and negatively correlated to $f_{\rm QPO}$ in the narrow frequency range of 1--1.8 Hz, then decreases slightly from $f_{\rm QPO}\sim2$ Hz to $f_{\rm QPO}\sim6$ Hz. Since the rms dependence on $f_{\rm QPO}$ of H 1743--322 in the rise phase is similar to that of GRS 1915+105, the $\delta \dot{H}_{\rm ext}$ dependence on $f_{\rm QPO}$ could also affect the rms-$f_{\rm QPO}$ relation of H 1743--322 in the rise phase. It may also work in the decay phase of H1743--322 if one considers in decay phase the positive and negative correlations at lower and higher frequencies, respectively. In Fig.~\ref{fig:6_p}, we show that the coronal temperature of the decay phase is higher than that of the rise phase, while there are no significant differences in photon index ($\Gamma$). In this case, as shown in Fig. 3 of \citet[][]{2022MNRAS.515.2099B}, the energy-averaged QPO rms in the energy range of 3--30 keV could be relatively lower in the decay phase. We propose the possible reason is that for the escaping photons in the specific energy range (3--30 keV), the higher coronal temperature indicates the less scatterings of the photons before escaping the Comptonizing medium, hence the effect of the variability amplitude amplification is weaker, if a balance between Compton cooling and external heating is at work for QPO amplification within the corona. We note that the above proposal is only qualitative, while the detailed investigation of the radiative properties of QPOs requires fitting the rms, phase lag, and time-averaged
spectra simultaneously with the time-dependent Comptonization model \citep[see][]{2021MNRAS.503.5522K,2022NatAs...6..577M,2022MNRAS.513.4196G,2022MNRAS.512.2686Z}.

\begin{figure}
\includegraphics[width=\linewidth]{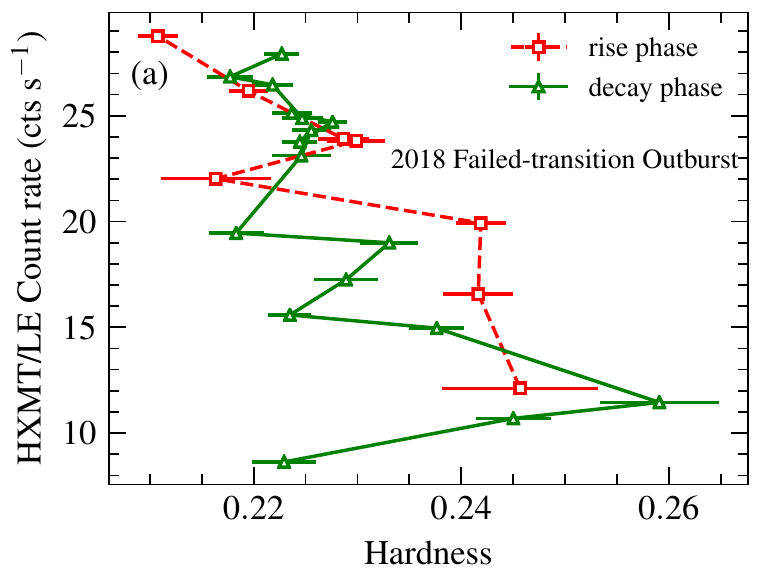}\\
\includegraphics[width=\linewidth]{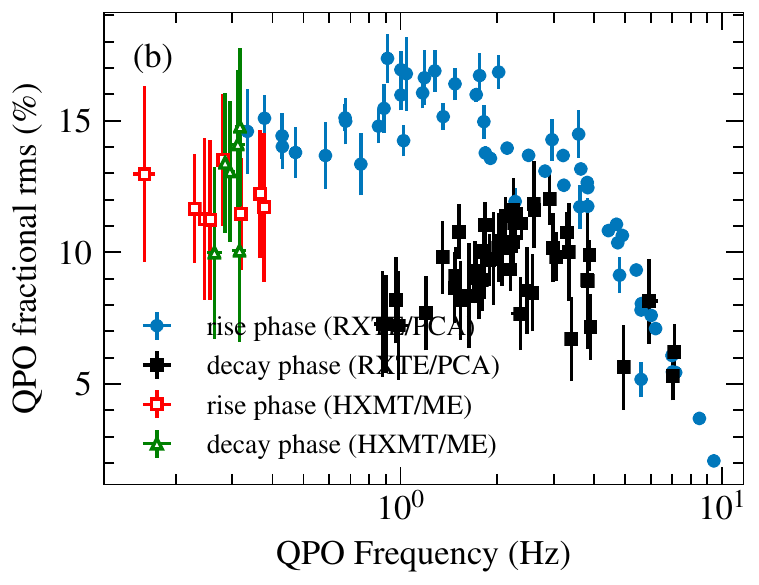}
\caption{The hardness-intensity diagram (HID) of the 2018 failed-transition outburst monitored by HXMT/LE (the top panel) and correlation between QPO rms and frequency using data from both of \emph{RXTE}/PCA and HXMT/ME (the bottom panel). The data of \emph{Insight}-HXMT are taken from \citet[][]{2022MNRAS.512.4541W}. The top panel displays the HID with separating the rise and decay phases: the rise data are plotted as red hollow squares, while the decay data are plotted as green hollow triangles. The bottom panel displays the correlations with separating data sets from different instruments and outburst phases.}
\label{fig:10}
\end{figure}

\subsection{Possible Relations Between Rms Differences and Hysteresis Effect}
The most recent failed-transition outburst of H 1743--322, monitored by \emph{Insight}-HXMT in 2018, has been reported by \citet{2022MNRAS.512.4541W}. Different from the outbursts presented above, this outburst remained in the LHS and never showed the significant hysteresis trace in the HID throughout the observed stage (see Fig.~\ref{fig:10}a). The \emph{Insight}-HXMT data of the 2018 outburst are taken from \citet{2022MNRAS.512.4541W}. We plot the fractional rms dependence on $f_{\rm QPO}$ using both \emph{Insight}-HXMT/ME and \emph{RXTE}/PCA data in Fig.~\ref{fig:10}b. As one can see, in the 2018 outburst, the QPO fractional rms remains roughly constant with a value  $\sim12\%$ without significant differences between the rise and decay phases. The 2008b outburst is also classified into failed-transition outbursts, however, the system experienced the LHS-HIMS transition and a hysteresis trace in the HID during this outburst \citep[see][]{2011MNRAS.414..677C}, with the rms differences between rise and decay phases which are consistent with those of other complete outbursts (see Fig.~\ref{fig:5}). On the basis of the above, the different rms-$f_{\rm QPO}$ relations between the rise and decay phases seem to be associated with the hysteresis trace in the HID, because these two phenomena accompany each other in H 1743--322. However, details about the relations between the rms differences and hysteresis effect need further investigations using more observational samples from more sources.

\subsection{Possible Scenario of the Corona-jet Coupling in H 1743--322}
Radio emission of BHXRBs is thought to be strongly related to relativistic jets \citep{2001MNRAS.322...31F}, which can also produce the Comptonization of soft photons from the disc \citep[see][]{1986ApJ...311..595B,2002A&A...388L..25G,2021A&A...646A.112R}. The jet is believed to be coupled to the accretion flow, but the nature of this connection is still not well understood. Based on a large dataset of \emph{RXTE} observations, frequent radio observations and the time-dependent Comptonization model, a series of studies have revealed a possible picture of the corona-jet coupling in GRS 1915+105 \citep[see][]{2020MNRAS.494.1375Z,2021MNRAS.503.5522K,2022NatAs...6..577M,2022MNRAS.513.4196G}. In H 1743--322, although quasi-simultaneous radio observations are not as abundant as that in GRS 1915+105, for each individual outburst, there is a marginally decreasing trend of the radio flux density from low QPO frequency ($<2$ Hz) to high QPO frequency ($\sim8$ Hz) in the rise phase (see Fig.~\ref{fig:6_p}), which indicates a quenching compact jet during the hard-to-soft state transition. \citet{2017MNRAS.464.2643V} found that the QPO hard phase lag became negative at high frequencies. Following the idea of the time-dependent Comptonization model \citep[see][]{2020MNRAS.492.1399K,2022MNRAS.515.2099B}, the nature of the hard lag (positive lag) is Comptonization, in which hard photons could experience more scatterings than soft ones before escaping the medium. The phase lag becomes negative due to the \emph{feedback} mechanism: the hard Comptonized photons may return back to the disc and be re-emitted later, so the softer photons could arrive later \citep[][]{1998MNRAS.299..479L}. Based on the above, we propose that the scenario of the corona-jet coupling in GRS 1915+105 can be applied to the rise phase of H 1743--322. At low QPO frequencies where the radio emission is strong, the Comptonizing medium is dominated by the jet-like corona, where the phase lag is positive \citep[see][]{2017MNRAS.464.2643V}. However, at high QPO frequencies where the radio emission is weak, the jet is quenched and replaced by an extended corona, which covers the inner parts of the thin disc. Since the disc has a large solid angle to receive the returning back photons, the \emph{feedback} effect could be strong enough to produce the observed negative phase lag \citep[][]{2017MNRAS.464.2643V}. We refer readers to \citet{2022NatAs...6..577M} for details about the physical picture of the coupling between the corona and the jet. In the outburst decay phase, the coronal temperature is relatively higher and the radio emission is weaker. If the corona is powered by the magnetic energy \citep[][]{2001MNRAS.321..549M}, especially as the case that the jet and the coronal power are tapped from the common magnetic energy reservoir \citep[see][]{2004MNRAS.351..253M}, the higher coronal temperature in the decay phase could be associated with the relatively weaker radio emission. On the basis of the above, we propose for H 1743--322 a similar scenario as the one proposed for GRS 1915+105 \citep[][]{2022NatAs...6..577M} which, as in that case, would account for the differences of the radio emission and coronal temperature between the two outburst stages. During the outburst decay phase, the magnetic field lines are disorganized and the magnetic energy is mainly dissipated stochastically in the corona, probably via magnetic reconnection, hence the jet is quenched and the coronal temperature is high. However, the magnetic configuration could be very different during the earlier stage of the rise phase ($f_{\rm QPO}<2$ Hz and the phase lag is positive), where the magnetic field lines are spatially coherent with a large scale poloidal component, which could channel materials out of the corona and collimate them in the direction perpendicular to the disc, then the Comptonization mainly occurs in the jet. During the hard-to-soft transition, the magnetic field lines vary from the coherent to the disorganized configuration, the jet is quenched and replaced by the extended corona. Such a different journey experienced by the accretion flow in the rise and decay phases may play a role in the difference seen in the rms-frequency relation (see Fig.~\ref{fig:5}).

\section{Conclusions}
\label{sec:5}
We performed systematic analyses of type C QPO observations from seven outbursts of H 1743--322 caught in the \emph{RXTE} era. 
With a number of observational samples, we confirm the independence of the type C QPO intrinsic rms on the individual outburst brightness which has been reported as well in GX 339--4. However, the dependence of QPO rms on frequency shows two branches in the outburst rise and decay phases, where the radio flux and coronal temperature are also different between the two phases. Both of the L-T precession model and the time-dependent Comptonization model can account for the rms difference, where the former needs a variable coronal geometric shape. Combining the recent \emph{Insight}-HXMT observations of this source during its failed-transition outburst, we suggest such the rms difference between the two outburst stages could be also related to the hysteresis effect in the HID. The co-evolution among the radio flux, coronal temperature and phase lags indicates there could be corona-jet transitions in H 1743--322 which have been recently reported in GRS 1915+105.

\begin{acknowledgments}
We are grateful to the anonymous referee for constructive comments that helped us improve this paper. This research has made use of data obtained from the High Energy Astrophysics Science Archive Research Center (HEASARC), provided by NASA’s Goddard Space Flight Center, and the \emph{Insight}-HXMT mission, a project funded by China National Space Administration (CNSA) and the Chinese Academy of Sciences (CAS). This work is supported by the National Key R\&D Program of China (2021YFA0718500) and the National Natural Science Foundation of China under grants, U1838201, U1838202, 12173103, U2038101 and U1938103. This work is partially supported by International Partnership Program of Chinese Academy of Sciences (Grant No.113111KYSB20190020).
\end{acknowledgments}
%






\appendix

\section{The distribution of reduced chi-square in the Spectral fitting}
\label{sec:A1}
\begin{figure}
\centering
	\includegraphics[width=8cm]{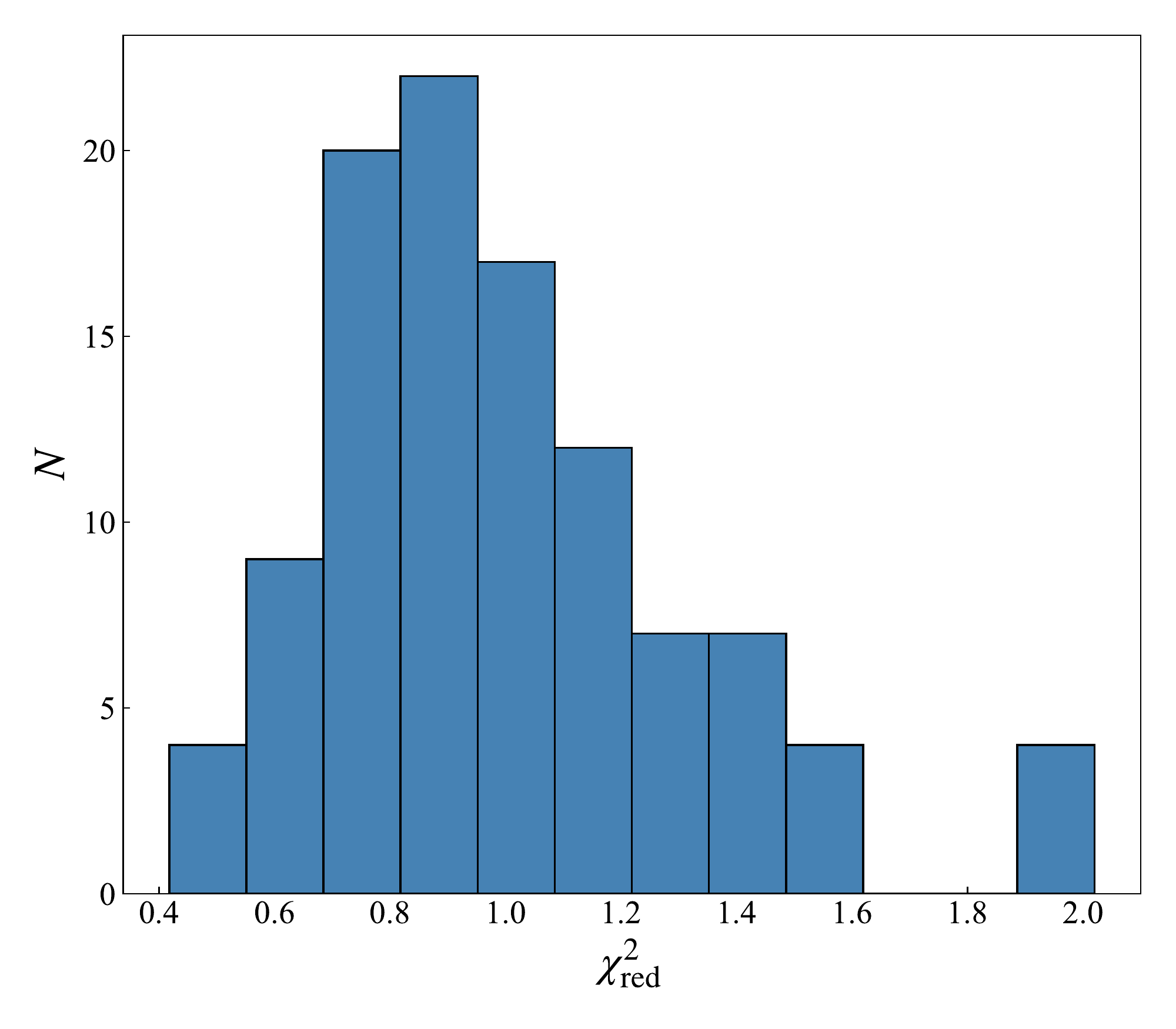}
    \caption{The reduced chi-square ($\chi^2_{\rm red}$) distribution of the spectral fitting with the Model 3 presented in Section~\ref{sec:3.2}.}
    \label{fig:A1}
\end{figure}

In this section, we show the reduced chi-square ($\chi^2_{\rm red}$) distribution of the spectral fitting with the Model 3 presented in Section~\ref{sec:3.2} (see Fig.~\ref{fig:A1}). There are one hundred and six observational samples in our spectral analysis.

\section{Calculation details of the coronal optical depth}
\label{sec:A2}
We consider a simplified coronal geometry with assuming the coronal shape is a crushed sphere (see Section~\ref{sec:3.4}). As a simplified model, the general relativity effects and Comptonization processes are neglected in our calculations. The effective optical depth of the corona is
\begin{equation}\label{eq:B1}
\tau=\alpha L,
\end{equation}
where $\alpha$ is the absorption coefficient, $L$ is the average size of the corona in the view of the observer. We can define the three-dimensional Cartesian coordinates $xyz$, where the $z$-axis is consistent with the normal of the corona ($\hat{\bm n}$), and the unit vector of the observer, $\hat{\bm o}$, is in the $yz$ plane. Then the ellipsoidal equation of the coronal surface is written by
\begin{equation}
\frac{x^2+y^2}{r^2}+\frac{z^2}{h^2}=1,
\end{equation}
where $r$ is the coronal radius and $h$ is the coronal height. We consider that the distance between the source and observer is far larger than the coronal radius (i.e. $D\gg r$), so the photon trajectories from the corona to the observer can be described as a cluster of parallel lines which are perpendicular to the $x$-axis. Then the linear equation of a representative photon trajectory to the observer is
\begin{align}
\label{eq:B3}
z&=\cot{\theta}\cdot y+z_0\notag\\
x&=x_0,
\end{align}
where $\theta$ is the include angle between $\hat{\bm n}$ and $\hat{\bm o}$, $z_0$ is the $z$-intercept of the projection of the line in the $yz$ plane, and $x_0$ is the $x$-intercept of the projection of the line in the $xy$ plane, respectively. Then the length of the corona which contributes to the flux in the direction of the representative photon trajectory is
\begin{equation}
l=\sqrt{1+\cot^2{\theta}}\frac{2hr\sqrt{\left(1-\frac{x_0^2}{r^2}\right)h^2+\left(1-\frac{x_0^2}{r^2}\right)r^2\cot^2{\theta}-z_0^2}}{h^2+r^2\cot^2{\theta}},
\end{equation}
i.e. the length of the line cut by the ellipsoidal surface. The average size of the corona in the view of the observer can be calculated by
\begin{align}\label{eq:B5}
L &= \frac{\int_{-r}^{r}\int_{-s}^{s}l(x_0,z_0) {\rm d}z_0{\rm d}x_0}{\int_{-r}^{r}\int_{-s}^{s}{\rm d}z_0{\rm d}x_0}\notag\\
&=\frac{4}{3}h\left\{\left(h/r\right)^2+\left[1-\left(h/r\right)^2\right]\cos^2{\theta}\right\}^{-1/2},
\end{align}
where $s\equiv\sqrt{\left[1-(x_0/r)^2\right]h^2+\left[1-(x_0/r)^2\right]r^2\cot^2{\theta}}$. When $z_0=\pm s$, the line of Eq.~\ref{eq:B3} is tangent to the ellipsoidal surface. Based on Eq.~\ref{eq:B1} and~\ref{eq:B5}, the optical depth can be written by
\begin{equation}
\tau=\frac{4}{3}\alpha h\left\{\left(h/r\right)^2+\left[1-\left(h/r\right)^2\right]\cos^2{\theta}\right\}^{-1/2}.
\end{equation}
We define $\tau_0\equiv\frac{4}{3}\alpha h$, where $\tau_0$ is the minimum optical depth of the corona, i.e. viewed from the coronal normal ($\theta=0$), then the optical depth can be written by
\begin{equation}
\tau=\tau_0\left\{\left(h/r\right)^2+\left[1-\left(h/r\right)^2\right]\cos^2{\theta}\right\}^{-1/2}.    
\end{equation}

\section{MCMC Parameter Probability Distributions}
\label{sec:A3}
\begin{figure*}
\begin{minipage}[c]{0.49\textwidth}
\centering
    \includegraphics[width=\linewidth]{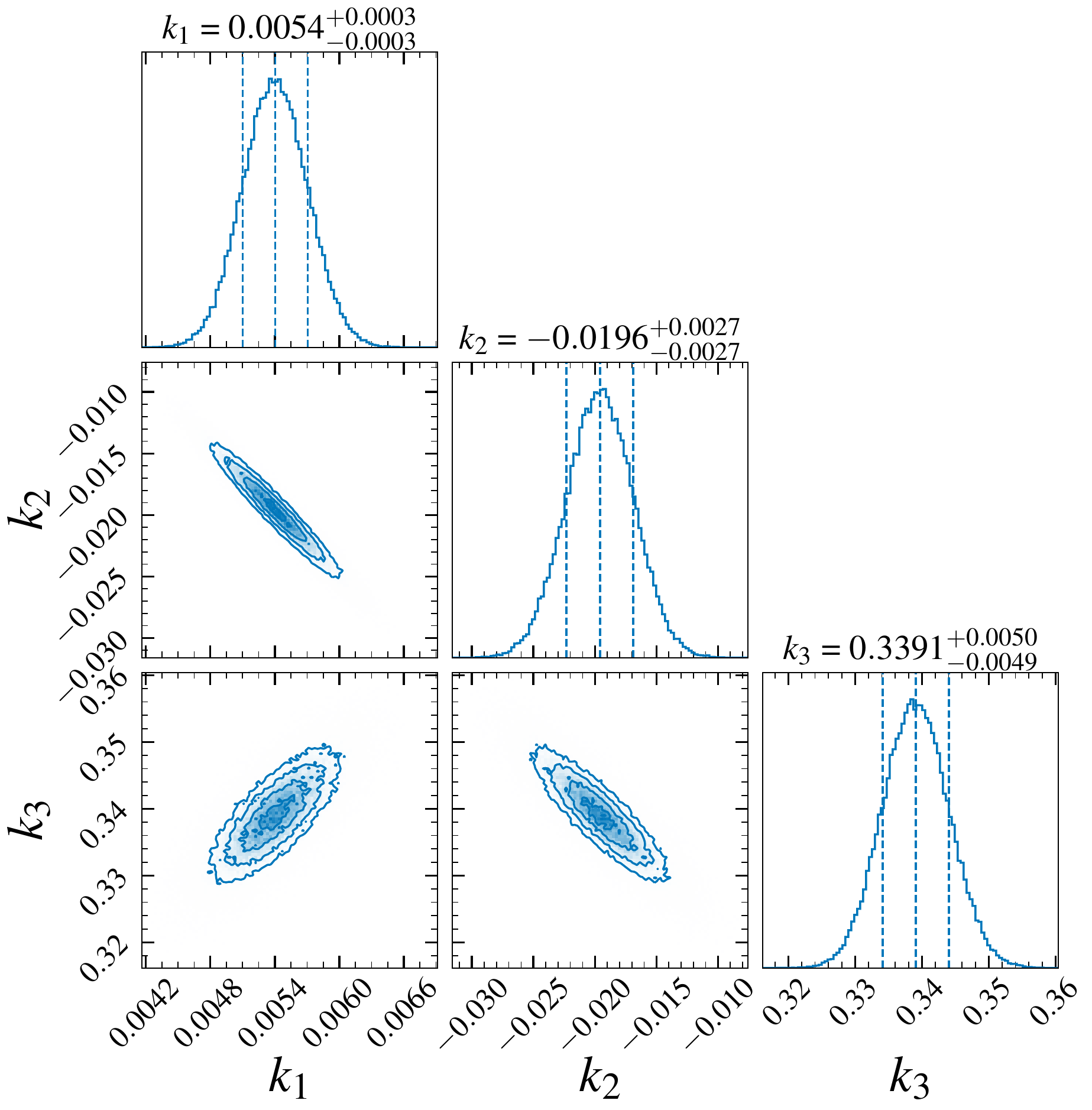}
\end{minipage}%
\begin{minipage}[c]{0.49\textwidth}
\centering
    \includegraphics[width=\linewidth]{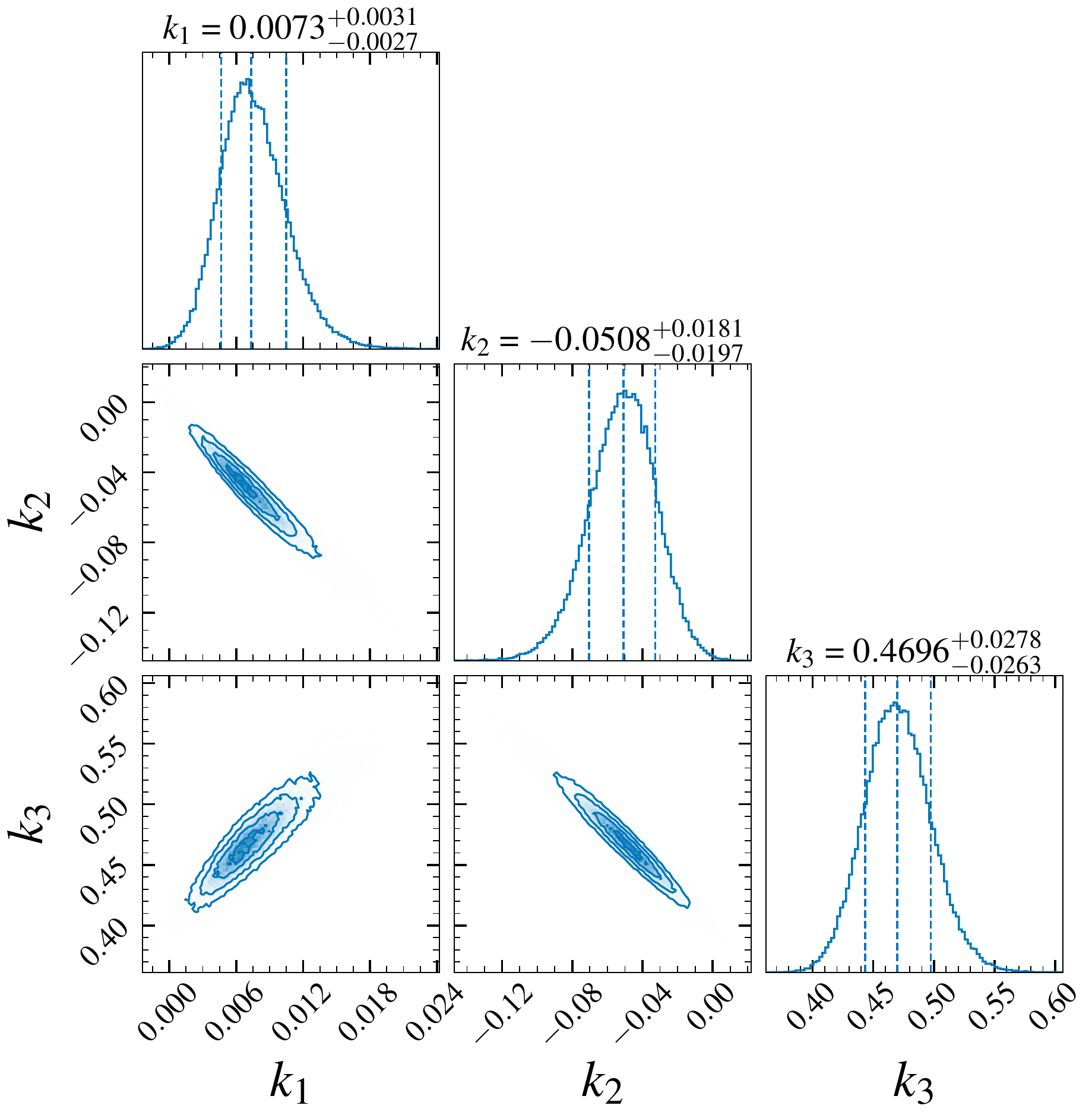}
\end{minipage}
    \caption{One- and two-dimensional projections of the posterior probability distributions, and the 0.16, 0.5 and 0.84 quantile contours derived from the MCMC analysis for the model parameters $k_1$, $k_2$ and $k_3$ described in Eq.~\ref{eq:13}. The left panel is plotted for the rise phase and the right panel is plotted for the decay phase.}
    \label{fig:A3}
\end{figure*}

In this section, we show the contour maps and probability distributions for the set of model parameters derived using the MCMC analysis of the relation between the QPO intrinsic rms and frequency. The MCMC analysis is preformed using \verb'emcee' package \citep{2013PASP..125..306F}, and the contour maps and probability distributions are plotted using \verb'corner' package \citep{2016JOSS....1...24F}. For each map, we show the 0.16, 0.5, and 0.84 quantiles (see Fig.~\ref{fig:A3}).

\section{QPO parameters and radio observations}
\label{sec:A4}
In this study, we preform a systematic analysis of type C QPOs with one hundred and six \emph{RXTE}/PCA observational samples of black hole X-ray binary H 1743--322. This source also exhibit significant radio emission, so we take the quasi-simultaneous radio observational results from previously published studies to add to our joint analysis. In this section, we present the QPO parameters of our analysis and the quasi-simultaneous radio flux measurements (if present) at $\sim 8.5$ GHz. The parameters, QPO frequency ($f_{\rm QPO}$), full width at half maximum (FWHM), fractional rms amplitude and radio flux density at $\sim 8.5$ GHz ($S_{\nu=8.5{\rm GHz}}$), etc., are presented in Table~\ref{tab:A4}.

\startlongtable
\begin{deluxetable}{lcccccccl}
\tablecaption{QPO Parameters and the Quasi-simultaneous Radio Flux Density\label{tab:A4}}
\tablewidth{\textwidth}
\tabletypesize{\scriptsize}
\tablehead{
\colhead{ObsID$^{\rm a}$} & Outburst Phase$^{\rm b}$ & \colhead{X-ray MJD$^{\rm c}$} &
\colhead{$f_{\rm QPO}$} & \colhead{FWHM$^{\rm d}$} &
\colhead{Fractional rms$^{\rm e}$} & \colhead{Radio MJD$^{\rm f}$} & \colhead{$S_{\nu=8.5{\rm GHz}}^{\rm g}$} & \colhead{Notes} \\ 
\colhead{} & \colhead{} & \colhead{} & \colhead{(Hz)} & \colhead{(Hz)} & 
\colhead{(\%)} & \colhead{} & \colhead{(mJy)} &
\colhead{}
} 
\startdata
{80138-01-06-00} & Rise & 52739.66 & $3.22_{-0.01}^{+0.01}$ & $0.35_{-0.02}^{+0.02}$ & $12.66_{-0.21}^{+0.22}$ & 52739.46 & $20.68\pm0.06$ &  VLA$^{\rm h}$ \\
{80138-01-07-00} & Rise & 52741.83 & $7.17_{-0.03}^{+0.03}$ & $1.13_{-0.07}^{+0.07}$ & $5.49_{-0.13}^{+0.13}$ & 52741.56 &  $7.71\pm0.12$ & VLA$^{\rm h}$ \\
{80146-01-01-00} & Rise & 52743.22 & $8.51_{-0.02}^{+0.02}$ & $0.56_{-0.04}^{+0.04}$ & $3.71_{-0.09}^{+0.09}$ & 52742.52 & $3.87\pm0.13$ & VLA$^{\rm h}$\\
{80146-01-02-00} & Rise & 52744.20 & $5.62_{-0.02}^{+0.02}$ & $0.85_{-0.04}^{+0.04}$ & $8.11_{-0.13}^{+0.13}$ & 52745.43 & $37.15\pm0.13$ & VLA$^{\rm h}$\\
{80146-01-03-00} & Rise & 52746.18 & $4.74_{-0.01}^{+0.01}$ & $0.55_{-0.03}^{+0.03}$ & $10.52_{-0.21}^{+0.22}$ & \nodata & \nodata & \nodata \\
{80146-01-03-01} & Rise & 52747.61 & $7.01_{-0.03}^{+0.03}$ & $1.19_{-0.08}^{+0.08}$ & $5.52_{-0.14}^{+0.14}$ & \nodata & \nodata & \nodata \\
{80146-01-29-00} & Rise & 52766.56 & $5.60_{-0.02}^{+0.02}$ & $0.74_{-0.04}^{+0.05}$ & $7.87_{-0.15}^{+0.15}$ & 52765.42 & $11.12\pm0.13$ & VLA$^{\rm h}$ \\
{80146-01-30-00} & Rise & 52767.81 & $4.43_{-0.01}^{+0.01}$ & $0.53_{-0.02}^{+0.02}$ & $10.92_{-0.13}^{+0.13}$ & 52767.51 & $23.02\pm0.12$ &  VLA$^{\rm h}$ \\
{80146-01-31-00} & Rise & 52768.53 & $5.41_{-0.02}^{+0.02}$ & $0.83_{-0.05}^{+0.05}$ & $9.41_{-0.18}^{+0.19}$ & 52768.49 & $30.14\pm0.16$ & VLA$^{\rm h}$ \\
{80146-01-32-00} & Rise & 52769.72 & $4.90_{-0.01}^{+0.01}$ & $0.55_{-0.03}^{+0.03}$ & $10.74_{-0.21}^{+0.23}$ & 52769.51 & $23.80\pm0.14$ & VLA$^{\rm h}$ \\
{80146-01-33-01} & Rise & 52770.37 & $6.02_{-0.03}^{+0.03}$ & $0.85_{-0.06}^{+0.07}$ & $7.66_{-0.20}^{+0.20}$ & \nodata & \nodata & \nodata \\
{80146-01-33-00} & Rise & 52770.65 & $6.22_{-0.02}^{+0.02}$ & $0.85_{-0.05}^{+0.06}$ & $7.16_{-0.15}^{+0.15}$ & \nodata & \nodata & \nodata \\
{80146-01-34-00} & Rise & 52771.74 & $2.82_{-0.01}^{+0.01}$ & $0.33_{-0.02}^{+0.02}$ & $13.22_{-0.17}^{+0.19}$ & \nodata & \nodata & \nodata \\
{80146-01-35-00} & Rise & 52771.97 & $2.27_{-0.01}^{+0.01}$ & $0.19_{-0.03}^{+0.03}$ & $12.06_{-0.52}^{+0.52}$ & \nodata & \nodata & \nodata \\
{80146-01-36-00} & Rise & 52772.67 & $1.84_{-0.01}^{+0.01}$ & $0.25_{-0.01}^{+0.01}$ & $13.92_{-0.26}^{+0.26}$ & \nodata & \nodata & \nodata \\
{80146-01-37-00} & Rise & 52773.66 & $1.90_{-0.01}^{+0.01}$ & $0.26_{-0.02}^{+0.02}$ & $13.71_{-0.25}^{+0.25}$ & 52773.36 & $35.76\pm0.23$ & VLA$^{\rm h}$ \\
{80146-01-39-00} & Rise & 52775.57 & $2.15_{-0.01}^{+0.01}$ & $0.29_{-0.02}^{+0.02}$ & $14.11_{-0.24}^{+0.24}$ & \nodata & \nodata & \nodata \\
{80146-01-40-00} & Rise & 52776.62 & $1.72_{-0.01}^{+0.01}$ & $0.32_{-0.02}^{+0.02}$ & $16.21_{-0.27}^{+0.27}$ & \nodata & \nodata & \nodata \\
{80146-01-41-00} & Rise & 52777.61 & $2.50_{-0.01}^{+0.01}$ & $0.31_{-0.02}^{+0.02}$ & $13.83_{-0.23}^{+0.22}$ & \nodata & \nodata & \nodata \\
{80146-01-42-00} & Rise & 52778.46 & $3.21_{-0.01}^{+0.01}$ & $0.46_{-0.02}^{+0.02}$ & $13.83_{-0.18}^{+0.19}$ & \nodata & \nodata & \nodata \\
{80146-01-43-01} & Rise & 52779.53 & $3.82_{-0.02}^{+0.02}$ & $0.41_{-0.03}^{+0.04}$ & $11.87_{-0.33}^{+0.34}$ & \nodata & \nodata & \nodata \\
{80146-01-43-00} & Rise & 52779.58 & $3.81_{-0.01}^{+0.01}$ & $0.37_{-0.02}^{+0.02}$ & $12.79_{-0.21}^{+0.22}$ & 52779.44 & $11.99\pm0.17$ & VLA$^{\rm h}$ \\
{80146-01-44-00} & Rise & 52780.57 & $3.82_{-0.01}^{+0.01}$ & $0.37_{-0.02}^{+0.02}$ & $12.59_{-0.21}^{+0.22}$ & \nodata & \nodata & \nodata \\
{80146-01-45-00} & Rise & 52781.55 & $3.64_{-0.01}^{+0.01}$ & $0.45_{-0.02}^{+0.02}$ & $13.31_{-0.17}^{+0.18}$ & \nodata & \nodata & \nodata \\
{80146-01-46-00} & Rise & 52782.67 & $4.70_{-0.01}^{+0.01}$ & $0.45_{-0.03}^{+0.03}$ & $11.18_{-0.19}^{+0.21}$ & \nodata & \nodata & \nodata \\
{80146-01-47-00} & Rise & 52783.46 & $6.99_{-0.02}^{+0.02}$ & $1.40_{-0.05}^{+0.06}$ & $6.13_{-0.10}^{+0.10}$ & \nodata & \nodata & \nodata \\
{80146-01-50-00} & Rise & 52786.29 & $9.43_{-0.04}^{+0.04}$ & $0.85_{-0.08}^{+0.09}$ & $2.09_{-0.08}^{+0.08}$ & 52786.36 & $16.06\pm0.11$ & VLA$^{\rm h}$ \\
{80137-01-25-00} & Decay & 52937.02 & $7.11_{-0.15}^{+0.21}$ & $0.91_{-0.41}^{+1.51}$ & $6.89_{-1.17}^{+1.18}$ & 52939.99 & $0.14\pm0.04$ & VLA$^{\rm h}$ \\
{80137-01-26-00} & Decay & 52938.00 & $5.93_{-0.39}^{+0.44}$ & $1.89_{-0.77}^{+1.32}$ & $9.05_{-1.66}^{+1.93}$ & \nodata & \nodata & \nodata \\
{80137-02-01-00} & Decay & 52944.11 & $2.48_{-0.11}^{+0.15}$ & $0.97_{-0.38}^{+0.68}$ & $10.03_{-1.75}^{+2.11}$ & \nodata & \nodata & \nodata \\
{93427-01-03-01} & Decay & 54492.18 & $7.03_{-0.10}^{+0.10}$ & $0.67_{-0.26}^{+0.41}$ & $5.81_{-0.87}^{+1.19}$ & 54493.32 & $0.44\pm0.09$ & ATCA$^{\rm i}$ \\
{93427-01-04-00} & Decay & 54498.84 & $3.90_{-0.08}^{+0.08}$ & $0.65_{-0.22}^{+0.33}$ & $8.00_{-1.10}^{+1.71}$ & 54499.74 & $0.52\pm0.06$ & VLA$^{\rm j}$\\
{93427-01-04-02} & Decay & 54500.80 & $2.60_{-0.10}^{+0.13}$ & $1.50_{-0.52}^{+0.82}$ & $13.42_{-1.77}^{+1.98}$ & 54501.64 & $0.48\pm0.08$ & VLA$^{\rm j}$ \\
{93427-01-04-03} & Decay & 54502.83 & $2.36_{-0.16}^{+0.14}$ & $0.77_{-0.35}^{+0.39}$ & $8.86_{-1.44}^{+1.74}$ & 54502.56 & $0.45\pm0.09$ & VLA$^{\rm j}$ \\
{93427-01-09-00} & Rise & 54742.98 & $0.33_{-0.01}^{+0.01}$ & $0.03_{-0.01}^{+0.01}$ & $15.19_{-1.67}^{+1.68}$ & 54744.21 & $1.74\pm0.07$ & ATCA$^{\rm i}$ \\
{93427-01-09-01} & Rise & 54746.51 & $0.38_{-0.01}^{+0.01}$ & $0.03_{-0.01}^{+0.01}$ & $15.69_{-0.89}^{+0.91}$ & \nodata & \nodata & \nodata \\
{93427-01-09-03} & Rise & 54747.49 & $0.43_{-0.01}^{+0.01}$ & $0.03_{-0.01}^{+0.01}$ & $15.00_{-0.89}^{+0.91}$ & 54747.44 & $2.54\pm0.08$ & ATCA$^{\rm i}$ \\
{93427-01-09-02} & Rise & 54748.21 & $0.47_{-0.01}^{+0.01}$ & $0.04_{-0.01}^{+0.01}$ & $14.31_{-0.98}^{+0.99}$ & 54748.44 & $2.43\pm0.09$ & ATCA$^{\rm i}$ \\
{93427-01-10-00} & Rise & 54750.37 & $0.58_{-0.01}^{+0.01}$ & $0.06_{-0.02}^{+0.02}$ & $14.19_{-1.31}^{+1.31}$ & 54749.36 & $2.38\pm0.11$ & ATCA$^{\rm i}$ \\
{93427-01-10-01} & Rise & 54752.26 & $0.68_{-0.01}^{+0.01}$ & $0.07_{-0.01}^{+0.01}$ & $15.59_{-0.91}^{+0.92}$ & \nodata & \nodata & \nodata \\
{93427-01-10-02} & Rise & 54755.23 & $0.75_{-0.02}^{+0.02}$ & $0.12_{-0.02}^{+0.03}$ & $13.88_{-1.24}^{+1.22}$ & \nodata & \nodata & \nodata \\
{93427-01-11-00} & Rise & 54756.25 & $0.86_{-0.01}^{+0.01}$ & $0.10_{-0.01}^{+0.01}$ & $15.44_{-0.67}^{+0.68}$ & \nodata & \nodata & \nodata \\
{93427-01-11-01} & Rise & 54758.15 & $1.02_{-0.01}^{+0.01}$ & $0.12_{-0.01}^{+0.02}$ & $14.89_{-0.62}^{+0.62}$ & \nodata & \nodata & \nodata \\
{93427-01-11-03} & Rise & 54762.14 & $5.60_{-0.06}^{+0.06}$ & $0.49_{-0.17}^{+0.23}$ & $5.43_{-0.66}^{+0.75}$ & \nodata & \nodata & \nodata \\
{93427-01-12-04} & Decay & 54767.84 & $3.87_{-0.03}^{+0.03}$ & $0.47_{-0.08}^{+0.09}$ & $10.48_{-0.62}^{+0.65}$ & \nodata & \nodata & \nodata \\
{93427-01-12-02} & Decay & 54768.64 & $3.30_{-0.03}^{+0.03}$ & $0.35_{-0.07}^{+0.09}$ & $11.45_{-0.81}^{+0.84}$ & \nodata & \nodata & \nodata \\
{93427-01-12-05} & Decay & 54769.14 & $2.91_{-0.04}^{+0.04}$ & $0.38_{-0.08}^{+0.10}$ & $12.88_{-1.05}^{+1.10}$ & \nodata & \nodata & \nodata \\
{93427-01-13-00} & Decay & 54770.12 & $2.61_{-0.02}^{+0.02}$ & $0.42_{-0.06}^{+0.06}$ & $12.40_{-0.61}^{+0.63}$ & \nodata & \nodata & \nodata \\
{93427-01-13-05} & Decay & 54770.40 & $2.26_{-0.02}^{+0.02}$ & $0.25_{-0.04}^{+0.05}$ & $11.90_{-0.68}^{+0.70}$ & \nodata & \nodata & \nodata \\
{93427-01-13-04} & Decay & 54771.76 & $2.38_{-0.02}^{+0.02}$ & $0.29_{-0.05}^{+0.05}$ & $11.89_{-0.68}^{+0.70}$ & \nodata & \nodata & \nodata \\
{93427-01-13-01} & Decay & 54772.15 & $2.25_{-0.03}^{+0.04}$ & $0.36_{-0.12}^{+0.15}$ & $12.44_{-1.27}^{+1.37}$ & \nodata & \nodata & \nodata \\
{93427-01-13-02} & Decay & 54773.27 & $2.27_{-0.03}^{+0.03}$ & $0.33_{-0.07}^{+0.08}$ & $12.21_{-0.86}^{+0.90}$ & \nodata & \nodata & \nodata \\
{93427-01-13-06} & Decay & 54774.64 & $2.12_{-0.02}^{+0.02}$ & $0.38_{-0.06}^{+0.07}$ & $12.26_{-0.75}^{+0.79}$ & 54774.43 & $0.94\pm0.12$ & ATCA$^{\rm i}$ \\
{93427-01-13-03} & Decay & 54775.56 & $1.83_{-0.03}^{+0.03}$ & $0.32_{-0.06}^{+0.07}$ & $11.92_{-0.92}^{+0.96}$ & \nodata & \nodata & \nodata \\
{93427-01-14-00} & Decay & 54777.86 & $1.89_{-0.04}^{+0.04}$ & $0.29_{-0.07}^{+0.08}$ & $10.59_{-1.04}^{+1.10}$ & \nodata & \nodata & \nodata \\
{93427-01-14-01} & Decay & 54778.77 & $1.82_{-0.03}^{+0.04}$ & $0.29_{-0.09}^{+0.12}$ & $10.99_{-1.21}^{+1.32}$ & \nodata & \nodata & \nodata \\
{93427-01-14-02} & Decay & 54779.04 & $1.52_{-0.02}^{+0.02}$ & $0.21_{-0.05}^{+0.06}$ & $11.68_{-1.09}^{+1.14}$ & 54779.35 & $0.94\pm0.08$ & ATCA$^{\rm i}$ \\
{93427-01-14-03} & Decay & 54780.02 & $1.48_{-0.03}^{+0.03}$ & $0.20_{-0.06}^{+0.07}$ & $9.89_{-1.16}^{+1.22}$ & \nodata & \nodata & \nodata \\
{93427-01-14-04} & Decay & 54781.78 & $1.72_{-0.02}^{+0.02}$ & $0.17_{-0.05}^{+0.07}$ & $9.88_{-1.01}^{+1.06}$ & \nodata & \nodata & \nodata \\
{93427-01-14-05} & Decay & 54782.89 & $2.21_{-0.04}^{+0.04}$ & $0.32_{-0.08}^{+0.10}$ & $11.15_{-1.11}^{+1.16}$ & \nodata & \nodata & \nodata \\
{93427-01-14-06} & Decay & 54783.81 & $2.07_{-0.05}^{+0.06}$ & $0.42_{-0.14}^{+0.20}$ & $11.16_{-1.44}^{+1.73}$ & \nodata & \nodata & \nodata \\
{93427-01-15-00} & Decay & 54784.45 & $1.79_{-0.04}^{+0.05}$ & $0.26_{-0.09}^{+0.12}$ & $9.74_{-1.22}^{+1.36}$ & \nodata & \nodata & \nodata \\
{93427-01-15-01} & Decay & 54785.70 & $1.95_{-0.08}^{+0.09}$ & $0.52_{-0.19}^{+0.27}$ & $10.83_{-1.74}^{+2.16}$ & \nodata & \nodata & \nodata \\
{93427-01-15-02} & Decay & 54786.48 & $1.76_{-0.03}^{+0.04}$ & $0.28_{-0.08}^{+0.11}$ & $9.31_{-1.07}^{+1.18}$ & \nodata & \nodata & \nodata \\
{93427-01-15-03} & Decay & 54787.73 & $1.54_{-0.06}^{+0.06}$ & $0.27_{-0.11}^{+0.16}$ & $9.15_{-1.57}^{+1.77}$ & \nodata & \nodata & \nodata \\
{93427-01-15-04} & Decay & 54788.64 & $0.97_{-0.03}^{+0.04}$ & $0.18_{-0.07}^{+0.14}$ & $9.29_{-1.67}^{+2.06}$ & \nodata & \nodata & \nodata \\
{93427-01-15-06} & Decay & 54788.84 & $0.99_{-0.07}^{+0.08}$ & $0.23_{-0.14}^{+0.19}$ & $8.09_{-2.08}^{+2.52}$ & \nodata & \nodata & \nodata \\
{93427-01-15-05} & Decay & 54789.49 & $0.88_{-0.05}^{+0.05}$ & $0.22_{-0.11}^{+0.24}$ & $8.20_{-1.82}^{+2.68}$ & \nodata & \nodata & \nodata \\
{94413-01-02-00} & Rise & 54980.40 & $0.91_{-0.01}^{+0.01}$ & $0.04_{-0.01}^{+0.01}$ & $17.96_{-0.96}^{+0.96}$ & 54978.38 & $2.24\pm0.03$ & VLA$^{\rm k}$ \\
{94413-01-02-02} & Rise & 54980.85 & $1.00_{-0.01}^{+0.01}$ & $0.03_{-0.01}^{+0.01}$ & $17.47_{-0.75}^{+0.75}$ & 54978.38 & $2.24\pm0.03$ & VLA$^{\rm k}$ \\
{94413-01-02-01} & Rise & 54981.95 & $1.19_{-0.01}^{+0.01}$ & $0.03_{-0.01}^{+0.01}$ & $17.16_{-1.07}^{+1.08}$ & 54981.95 & $2.73\pm0.10$ & VLA$^{\rm k}$ \\
{94413-01-02-05} & Rise & 54982.28 & $1.28_{-0.01}^{+0.01}$ & $0.04_{-0.01}^{+0.01}$ & $17.47_{-0.81}^{+0.82}$ & \nodata & \nodata & \nodata \\
{94413-01-02-04} & Rise & 54983.33 & $2.02_{-0.01}^{+0.01}$ & $0.09_{-0.01}^{+0.01}$ & $17.43_{-0.66}^{+0.67}$ & \nodata & \nodata & \nodata \\
{94413-01-02-03} & Rise & 54984.37 & $3.58_{-0.01}^{+0.01}$ & $0.33_{-0.04}^{+0.04}$ & $14.97_{-0.69}^{+1.21}$ & 54984.35 & $1.8\pm0.3$ & VLBA$^{\rm l}$ \\
{94413-01-07-00} & Decay & 55016.32 & $4.94_{-0.12}^{+0.06}$ & $0.26_{-0.26}^{+0.45}$ & $6.22_{-1.46}^{+2.10}$ & \nodata & \nodata & \nodata \\
{94413-01-07-01} & Decay & 55019.45 & $3.40_{-0.10}^{+0.09}$ & $0.39_{-0.18}^{+0.29}$ & $7.49_{-1.54}^{+2.04}$ & 55019.46 & $0.592\pm0.055$ & VLA$^{\rm k}$ \\
{94413-01-07-02} & Decay & 55021.42 & $3.81_{-0.11}^{+0.17}$ & $0.62_{-0.29}^{+0.85}$ & $10.16_{-2.10}^{+3.87}$ & 55021.42 & $0.410\pm0.074$ & VLA$^{\rm k}$ \\
{95405-01-02-06} & Decay & 55223.39 & $3.82_{-0.04}^{+0.04}$ & $0.43_{-0.10}^{+0.13}$ & $9.55_{-0.84}^{+0.91}$ & \nodata & \nodata & \nodata \\
{95405-01-02-02} & Decay & 55224.67 & $3.05_{-0.04}^{+0.04}$ & $0.39_{-0.09}^{+0.12}$ & $10.54_{-0.95}^{+1.04}$ & \nodata & \nodata & \nodata \\
{95405-01-03-00} & Decay & 55226.53 & $2.02_{-0.03}^{+0.04}$ & $0.30_{-0.08}^{+0.10}$ & $11.23_{-1.11}^{+1.19}$ & \nodata & \nodata & \nodata \\
{95405-01-03-04} & Decay & 55227.77 & $2.24_{-0.04}^{+0.03}$ & $0.30_{-0.07}^{+0.09}$ & $11.36_{-1.07}^{+1.12}$ & \nodata & \nodata & \nodata \\
{95405-01-03-01} & Decay & 55228.61 & $2.20_{-0.03}^{+0.03}$ & $0.26_{-0.06}^{+0.07}$ & $10.21_{-0.86}^{+0.89}$ & \nodata & \nodata & \nodata \\
{95405-01-03-05} & Decay & 55229.54 & $1.71_{-0.04}^{+0.04}$ & $0.24_{-0.07}^{+0.09}$ & $10.10_{-1.22}^{+1.31}$ & \nodata & \nodata & \nodata \\
{95405-01-03-02} & Decay & 55230.59 & $1.35_{-0.04}^{+0.04}$ & $0.32_{-0.10}^{+0.14}$ & $10.75_{-1.40}^{+1.62}$ & \nodata & \nodata & \nodata \\
{95405-01-04-01} & Decay & 55233.40 & $0.90_{-0.03}^{+0.03}$ & $0.15_{-0.09}^{+0.17}$ & $8.15_{-1.69}^{+2.44}$ & \nodata & \nodata & \nodata \\
{95360-14-01-00} & Rise & 55418.41 & $1.00_{-0.01}^{+0.01}$ & $0.05_{-0.01}^{+0.01}$ & $16.56_{-0.60}^{+0.60}$ & \nodata & \nodata & \nodata \\
{95360-14-02-01} & Rise & 55419.09 & $1.04_{-0.01}^{+0.01}$ & $0.04_{-0.01}^{+0.01}$ & $17.38_{-1.45}^{+1.46}$ & \nodata & \nodata & \nodata \\
{95360-14-02-00} & Rise & 55420.23 & $1.17_{-0.01}^{+0.01}$ & $0.04_{-0.01}^{+0.01}$ & $16.66_{-0.61}^{+0.61}$ & \nodata & \nodata & \nodata \\
{95360-14-03-00} & Rise & 55421.28 & $1.48_{-0.01}^{+0.01}$ & $0.05_{-0.01}^{+0.01}$ & $17.02_{-0.64}^{+0.64}$ & \nodata & \nodata & \nodata \\
{95360-14-02-03} & Rise & 55422.02 & $1.76_{-0.01}^{+0.01}$ & $0.08_{-0.01}^{+0.01}$ & $17.30_{-0.89}^{+0.89}$ & \nodata & \nodata & \nodata \\
{95360-14-02-02} & Rise & 55423.20 & $2.96_{-0.01}^{+0.02}$ & $0.20_{-0.03}^{+0.04}$ & $14.88_{-0.80}^{+0.87}$ & \nodata & \nodata & \nodata \\
{95360-14-03-01} & Rise & 55424.06 & $4.80_{-0.04}^{+0.04}$ & $0.37_{-0.07}^{+0.09}$ & $9.44_{-0.70}^{+0.71}$ & \nodata & \nodata & \nodata \\
{95360-14-23-00} & Decay & 55456.67 & $3.34_{-0.07}^{+0.08}$ & $0.58_{-0.22}^{+0.38}$ & $11.00_{-1.77}^{+2.30}$ & \nodata & \nodata & \nodata \\
{95360-14-23-01} & Decay & 55457.12 & $2.57_{-0.04}^{+0.05}$ & $0.25_{-0.11}^{+0.18}$ & $9.34_{-1.47}^{+1.77}$ & \nodata & \nodata & \nodata \\
{96425-01-01-00} & Rise & 55663.67 & $0.43_{-0.01}^{+0.01}$ & $0.03_{-0.01}^{+0.01}$ & $14.61_{-0.87}^{+0.88}$ & \nodata & \nodata & \nodata \\
{96425-01-02-00} & Rise & 55667.59 & $0.67_{-0.01}^{+0.01}$ & $0.05_{-0.01}^{+0.01}$ & $15.72_{-0.53}^{+0.54}$ & \nodata & \nodata & \nodata \\
{96425-01-02-01} & Rise & 55668.98 & $0.89_{-0.01}^{+0.01}$ & $0.08_{-0.01}^{+0.02}$ & $16.05_{-0.96}^{+0.99}$ & \nodata & \nodata & \nodata \\
{96425-01-02-02} & Rise & 55670.62 & $1.36_{-0.01}^{+0.01}$ & $0.14_{-0.01}^{+0.01}$ & $15.80_{-0.52}^{+0.53}$ & \nodata & \nodata & \nodata \\
{96425-01-02-05} & Rise & 55671.53 & $1.82_{-0.01}^{+0.01}$ & $0.21_{-0.03}^{+0.03}$ & $15.61_{-0.67}^{+0.67}$ & \nodata & \nodata & \nodata \\
{96425-01-02-03} & Rise & 55672.84 & $3.61_{-0.03}^{+0.03}$ & $0.29_{-0.06}^{+0.07}$ & $12.21_{-0.86}^{+0.87}$ & \nodata & \nodata & \nodata \\
{96425-01-05-01} & Decay & 55690.13 & $2.98_{-0.07}^{+0.07}$ & $0.54_{-0.17}^{+0.27}$ & $10.99_{-1.35}^{+1.49}$ & \nodata & \nodata & \nodata \\
{96425-01-05-02} & Decay & 55691.51 & $2.01_{-0.03}^{+0.03}$ & $0.24_{-0.11}^{+0.16}$ & $11.00_{-1.31}^{+1.51}$ & \nodata & \nodata & \nodata \\
{96425-01-05-03} & Decay & 55693.00 & $1.81_{-0.04}^{+0.04}$ & $0.28_{-0.10}^{+0.14}$ & $10.88_{-1.34}^{+1.47}$ & \nodata & \nodata & \nodata \\
{96425-01-06-00} & Decay & 55694.00 & $1.47_{-0.03}^{+0.04}$ & $0.24_{-0.11}^{+0.16}$ & $9.48_{-1.39}^{+1.59}$ & \nodata & \nodata & \nodata \\
{96425-01-06-01} & Decay & 55695.42 & $1.20_{-0.04}^{+0.04}$ & $0.19_{-0.08}^{+0.14}$ & $8.43_{-1.47}^{+1.64}$ & \nodata & \nodata & \nodata \\
\enddata
\tablecomments{$^{\rm a}$ Observational ID of \emph{RXTE} data. $^{\rm b}$ Outburst phase (rise phase or decay phase). $^{\rm c}$ Modified Julian Day of the \emph{RXTE} observation. $^{\rm d}$ Full width at half maximum of the QPO. $^{\rm e}$ The QPO fractional rms calculated by Eq.~\ref{eq:1}. $^{\rm f}$ Modified Julian Day of the radio observation. $^{\rm g}$ The radio flux density at $\sim$8.5 GHz. $^{\rm h}$ The radio flux measurements of Very Large Array (VLA) are taken from \citet{2009ApJ...698.1398M}. $^{\rm i}$ The radio flux measurements of Australia Telescope Compact Array (ATCA) are taken from \citet{2011MNRAS.414..677C}. $^{\rm j}$ The radio flux measurements of Very Large Array (VLA) are taken from \citet{2010MNRAS.401.1255J}. $^{\rm k}$ The radio flux measurements of Very Large Array (VLA) are taken from \citet{2011MNRAS.414..677C}. $^{\rm l}$ The radio flux measurements of Very Long Baseline Array (VLBA) are taken from \citet{2012MNRAS.421..468M}.}
\end{deluxetable}


\bibliography{sample631}{}
\bibliographystyle{aasjournal}



\end{document}